\documentclass{jpp}
\usepackage{graphicx}                       
\usepackage[outdir=./]{epstopdf}
\usepackage{amsmath}
\usepackage{amssymb}
\usepackage{bbold}
\usepackage[usenames,dvipsnames,table]{xcolor}
\usepackage[format=plain, font=small, labelfont=bf, justification = raggedright]{caption}
\usepackage{array}  
\usepackage{overpic}
\usepackage{placeins} 
\usepackage{fancyhdr}
\usepackage{empheq} 
\usepackage[makeroom]{cancel} 
\usepackage{multirow} 
\usepackage{hyperref} 
\hypersetup{colorlinks = true,linkcolor = blue, breaklinks = true}

\newcommand{\eq}[1]{(\ref{#1})}
\newcommand{\Eq}[1]{Eq.~\eq{#1}}
\newcommand{\Eqs}[1]{Eqs.~\eq{#1}}
\newcommand{\Fig}[1]{Fig.~\ref{#1}}
\newcommand{\Sec}[1]{Sec.~\ref{#1}}

\newcommand{\App}[1]{Appendix~\ref{#1}}

\newcommand{\ie}{{i.e., }}

\newcommand{\mc}[1]{\mathcal{#1}}

\newcommand{\Vect}[1]{{\boldsymbol{\rm #1}}}

\newcommand{\fourier}[1]{\smash{\widetilde{#1}}}

\newcommand{\pd}[1]{\partial_{#1}}
\newcommand{\dd}{\mathrm{d}}

\DeclareMathOperator{\airyA}{Ai}
\DeclareMathOperator{\airyB}{Bi}
\DeclareMathOperator{\airyG}{Gi}

\renewcommand{\Re}{\textrm{Re}}
\renewcommand{\Im}{\textrm{Im}}

\newcommand{\airyLEN}{\ell}
\newcommand{\airySKIN}{\delta_a}
\newcommand{\rppLOC}{d}
\newcommand{\width}{W}
\newcommand{\compFOCALnorm}{q_c}
\newcommand{\norm}{\mc{E}}

\newcommand{\normLKS}{\mc{E}_\text{L}}

\newcommand{\numRPP}{M}
\newcommand{\enumRPP}{m}
\newcommand{\thetaRPP}{\phi_{\enumRPP}}

\newcommand{\coupling}{\eta}
\newcommand{\fnum}{f_\#}

\newcommand{\Stroke}[1]{\text{\ooalign{ $#1$\cr \hidewidth\raise.225ex \hbox{$-\mkern.5mu$}\cr}}}

\graphicspath{{/Users/nicolaslopez/Desktop/Figures/Exact_Airy/}}

\begin{document}
\setlength{\parskip}{0pt}
\setlength{\belowcaptionskip}{0pt}


\shorttitle{Exact boundary-value solution to linear layer problem}
\shortauthor{N.~A.~Lopez}

\title{Exact boundary-value solution for an electromagnetic wave propagating in a linearly-varying index of refraction}
\author{N.~A.~Lopez\aff{1}
}

\affiliation{\aff{1}Rudolf Peierls Centre for Theoretical Physics, University of Oxford,
Oxford OX1 3PU, UK}

\maketitle

\begin{abstract}
    The propagation of electromagnetic waves in a linearly-varying index of refraction is a fundamental problem in wave physics, being relevant in fusion science for describing certain wave-based heating and diagnostic schemes. Here, an exact solution is obtained for a given incoming wavefield specified on the boundary transverse to the direction of inhomogeneity by performing a spectral, rather than asymptotic, matching. Two case studies are then presented: a Gaussian beam at oblique incidence and a speckled wavefield at normal incidence. For the Gaussian beam, it is shown that when the waist $W$ is sufficiently large, oblique incidence manifests simply as rigid translation and focal shift of the corresponding diffraction pattern at normal incidence. The destruction of the hyperbolic umbilic caustic (corresponding to a critically focused beam) as $W$ is reduced is then demonstrated. The caustic disappears once $W \lesssim \airySKIN \sqrt{L}$ ($L$ being the medium lengthscale normalized by the Airy skin depth $\airySKIN$), at which point the wave behavior is increasingly described by Airy functions, but experiences less focusing as a result. To maximize the intensity of a launched Gaussian beam at a turning point, one should therefore minimize the imaginary part of the launched complex beam parameter while having the real part satisfy critical focusing. For the speckled wavefield, it is shown that the transverse speckle pattern only couples to the Airy longitudinal pattern when the coupling parameter $\coupling = \sqrt{L}/\fnum$ is large, with $\fnum$ being the f-number of the launching aperture. When $\coupling \ll 1$, a reduced description of the total wavefield can be obtained by simply multiplying the incoming speckle pattern with the Airy swelling.
\end{abstract}

\section{Introduction}

The description of an electromagnetic wave propagating in a linearly-varying medium is a classic problem in wave physics, sometimes referred to as the linear-layer problem~\citep{Ginzburg61}. Its relevance to plasma physics and fusion research is predominantly as a local description near a turning point for wave-based heating and diagnostics applications. For example, a recently developed scheme to perform fundamental X-mode heating and current drive in startup plasmas~\citep{Ono22a,Ono22b} involves an obliquely launched X-mode resonating with electrons near the low-density cutoff; the efficiency of the heating process will therefore depend on the wavefield intensity behavior near the cutoff. Similarly, the Doppler backscattering diagnostic~\citep{HallChen22} relies on the swelling of a wavefield near a turning point to probe turbulence spectra via nonlinear scattering of the diagnostic beam; the intensity details near the turning point therefore determine the diagnostic sensitivity and localization.

Many authors~\citep{Ginzburg61,Orlov80,Maj09,Maj10,Lopez23} have obtained solutions to the linear-layer problem in terms of a prescribed field $\psi$ along the boundary $z = 0$, where $z$ is the direction of medium inhomogeneity. However, the full field $\psi$ is made up of both the incoming and reflected components, and often only the incoming component is known in practice. To remedy this apparent shortcoming in the obtained solutions, most of the aforementioned authors performed an asymptotic matching to isolate only the incoming component. This resolves the issue with the boundary condition, but at the cost of introducing an asymptotic validity criterion into the analysis: the resulting formulas are not valid when the incoming field is specified too close to the turning point, as might occur when beams are launched obliquely. A universally valid formula that matches to a prescribed incoming field would be more desirable for use in applications.

Here we obtain such a formula by performing the matching spectrally instead of asymptotically. No new asymptotic validity criteria are introduced into the problem, and it is shown explicitly that the obtained solution exactly reproduces the incoming wavefield at the boundary, regardless the distance between the boundary and the turning point. The solution involves an integral whose kernel contains the standard Airy function, which is expected to arise in this problem, and also the related Scorer function, which is less well-known. To demonstrate the flexibility of the new solution, two special cases are considered: an obliquely launched Gaussian beam and a normally incident speckled wavefield produced by a bilevel random phase plate (RPP). For this first example, it is shown how the hyperbolic umbilic caustic created at critical focusing gets degraded as the beam waist becomes smaller for a variety of injection angles, including angles at which the asymptotic validity criterion for the previous solutions is violated. For the second example, an explicit coupling parameter is derived and demonstrated that governs whether the speckles influence the behavior of the wavefield near the turning point. Both these examples can serve as starting points to developing reduced models of waves near turning points with more comprehensive physics content, which would be useful for the fusion applications mentioned in the first paragraph, among other applications.

This paper is organized as follows. In \Sec{sec:background} the linear-layer problem is introduced. In \Sec{sec:match} a spectral matching is performed to allow the solution to the linear-layer problem to be expressed only in terms of the incoming field at the boundary. This is the main result of this paper. In \Sec{sec:gauss} the special case of an incoming Gaussian beam is studied, with particular emphasis on its behavior near critical focusing. In \Sec{sec:speckle} the special case of an incoming speckled wavefield is studied, with focus on characterizing the coupling between speckles and the Airy pattern. Finally, \Sec{sec:concl} summarizes the main conclusions. Auxiliary calculations are presented in appendices.

\section{Background}
\label{sec:background}

Let us consider an electromagnetic wave propagating in $N + 1$ spatial dimensions in a medium whose index of refraction varies as a linear function. We take one dimension, denoted $z$, to be aligned with the direction of inhomogeneity, and label the remaining $N$ dimensions by the vector $\Vect{x}$. Assuming time-harmonic modes with a single angular frequency $\omega$, the wavefield amplitude can be shown to satisfy the Helmholtz equation
\begin{equation}
	\pd{\Vect{x}}^2 \psi(\Vect{x},z) + \pd{z}^2 \psi(\Vect{x},z) + \frac{\airyLEN - z}{\airySKIN^3} \psi(\Vect{x},z) = 0,
    \label{eq:airySTART}
\end{equation}

\noindent where $\airySKIN$ is a constant with units of length sometimes called the `Airy skin depth'~\citep{Michel23}. In terms of the angular frequency $\omega$, the medium lengthscale $\airyLEN$, and the speed of light in vacuum $c$, $\airySKIN$ is given as
\begin{equation}
    \airySKIN = \sqrt[3]{
    \frac{\airyLEN c^2 }{\omega^2}
    }
    .
    \label{eq:skinDEF}
\end{equation}

\noindent We shall maintain $N$ unspecified in the following analysis, but note that practical calculations will have either $N = 1$ or $N = 2$ for $2$-D or $3$-D propagation, respectively. We shall also only seek solutions that are stable, such that they become evanescent and decay to zero as $z \to +\infty$ (\ie we shall impose a radiation boundary condition).

Let us now introduce normalized spatial coordinates
\begin{equation}
    \Vect{x} = \airySKIN \Vect{X}
	, \quad z = \airySKIN Z
	, \quad \airyLEN = \airySKIN L
    ,
\end{equation}

\noindent such that \Eq{eq:airySTART} becomes
\begin{equation}
    \pd{\Vect{X}}^2 \psi(\Vect{X},Z) + \pd{Z}^2 \psi(\Vect{X},Z) + (L - Z) \psi(\Vect{X},Z) = 0 .
    \label{eq:airyNORMALIZED}
\end{equation}

\noindent Let us also adopt the following convention for the Fourier transform (FT):
\begin{subequations}
    \label{eq:FT}
    \begin{align}
        \fourier{\psi}(\Vect{K}_x,K_z) &= 
        \int \frac{\dd \Vect{X} \, \dd Z}{(2\pi)^{N + 1}}
    	\psi(\Vect{X},Z) e^{- i \Vect{K}_x \cdot \Vect{X} - i K_z Z}
        , \\
        \psi(\Vect{X},Z) &= 
        \int \dd \Vect{K}_x \, \dd K_z \,
        \fourier{\psi}(\Vect{K}_x,K_z) e^{i \Vect{K}_x \cdot \Vect{X} + i K_z Z}
        .
    \end{align}
\end{subequations}

\noindent Taking the FT of \Eq{eq:airyNORMALIZED} then gives
\begin{equation}
	i \pd{K_z} \fourier{\psi}(\Vect{K}_x,K_z)
	= \left(
		L
		- |\Vect{K}_x|^2
		- K_z^2
	\right) \fourier{\psi}(\Vect{K}_x,K_z)
    ,
\end{equation}

\noindent with solution given by
\begin{equation}
	\fourier{\psi}(\Vect{K}_x,K_z)
	= \fourier{\psi}_0(\Vect{K}_x)
	\exp\left[
		i \frac{K_z^3}{3}
		+ i \left( |\Vect{K}_x|^2 - L \right) K_z
	\right]
    .
    \label{eq:FTsol}
\end{equation}

\noindent Here, $\fourier{\psi}_0(\Vect{K}_x) \equiv \fourier{\psi}(0,\Vect{K}_x)$ is an arbitrary function that can eventually be matched to boundary conditions, as we show in the next section. 

The general solution to \Eq{eq:airyNORMALIZED} is then obtained by taking an inverse FT of \Eq{eq:FTsol}:
\begin{align}
	&\psi(\Vect{X},Z)
	= \int \dd \Vect{K}_x \, \dd K_z \,
    \fourier{\psi}_0(\Vect{K}_x)
	\exp\left[
		i \frac{K_z^3}{3}
		+ i \left( |\Vect{K}_x|^2 + Z - L \right) K_z
        + i \Vect{K}_x \cdot \Vect{X}
	\right]
    .
    \label{eq:solGENERAL}
\end{align}

\noindent Note that the radiation boundary condition has been tacitly imposed through our use of an FT to obtain the solution \eq{eq:solGENERAL}.

\section{Boundary-value solution for prescribed incoming wavefield}
\label{sec:match}

\subsection{Isolating the incoming contribution}

Although \Eq{eq:solGENERAL} constitutes the general boundary-value solution for a prescribed $\fourier{\psi}_0(\Vect{K}_x)$, this boundary-value solution is formulated in the spectral domain, which is not useful for most applications. Here we shall instead obtain the boundary-value solution in the coordinate domain by prescribing an incoming wavefield, since this is often what is known in practice.

To do this, consider the field on the $Z = 0$ plane, \ie $\psi(\Vect{X},0)$. We can split this into `incoming' and `outgoing' components by using a spectral filter based on the sign of $K_z$ as follows:
\begin{subequations}
    \begin{align}
        \hspace{-4mm}\psi_\text{in}(\Vect{X}) &=
        \int \dd \Vect{K}_x
        \, \fourier{\psi}_0(\Vect{K}_x) e^{i \Vect{K}_x \cdot \Vect{X}}
        \int_0^\infty \dd K_z \,
        \exp\left[
            i \frac{K_z^3}{3}
            + i \left( |\Vect{K}_x|^2 - L \right) K_z
        \right]
        , \\
        \hspace{-4mm}\psi_\text{out}(\Vect{X}) &=
        \int \dd \Vect{K}_x
        \, \fourier{\psi}_0(\Vect{K}_x) e^{i \Vect{K}_x \cdot \Vect{X}}
        \int_{-\infty}^0 \dd K_z \,
        \exp\left[
            i \frac{K_z^3}{3}
            + i \left( |\Vect{K}_x|^2 - L \right) K_z
        \right]
        .
    \end{align}
\end{subequations}

\noindent One then has the exact decomposition
\begin{equation}
    \psi(\Vect{X},0) =
	\psi_\text{in}(\Vect{X})
	+ \psi_\text{out}(\Vect{X})
    .
\end{equation}

To proceed further, one is required to compute the integral
\begin{align}
	I(\zeta) 
    &= \int_0^\infty \dd K_z \,
	\exp\left(
		i \frac{K_z^3}{3}
		+ i \zeta K_z
	\right)
    \nonumber\\
	&\equiv \int_0^\infty \dd K_z \,
	\cos\left(
		\frac{K_z^3}{3}
		+ \zeta K_z
	\right)
	+ i \int_0^\infty \dd K_z \, 
	\sin\left(
		\frac{K_z^3}{3}
		+ \zeta K_z
	\right)
    ,
    \label{eq:intREQ}
\end{align}

\noindent where $\zeta$ is a real-valued parameter. Note that the other relevant integral is
\begin{equation}
    \int_{-\infty}^0 \dd K_z \, 
	\exp\left[
		i \frac{K_z^3}{3}
		+ i \zeta K_z
	\right]
    = \left[ I(\zeta) \right]^*
    .
\end{equation}

\noindent Both the integrals involved in the real and imaginary parts of \Eq{eq:intREQ} can be solved in terms of Airy-related functions~\citep{Olver10a}:
\begin{align}
    \int_0^\infty \dd K_z \, \cos \left( \frac{K_z^3}{3} + \zeta K_z \right) = \pi \airyA(\zeta)
    , \quad
    \int_0^\infty \dd K_z \, \sin \left( \frac{K_z^3}{3} + \zeta K_z \right) = \pi \airyG(\zeta)
    ,
\end{align}

\begin{figure}
	\centering\includegraphics[width=0.6\linewidth,trim={2mm 4mm 2mm 2mm},clip]{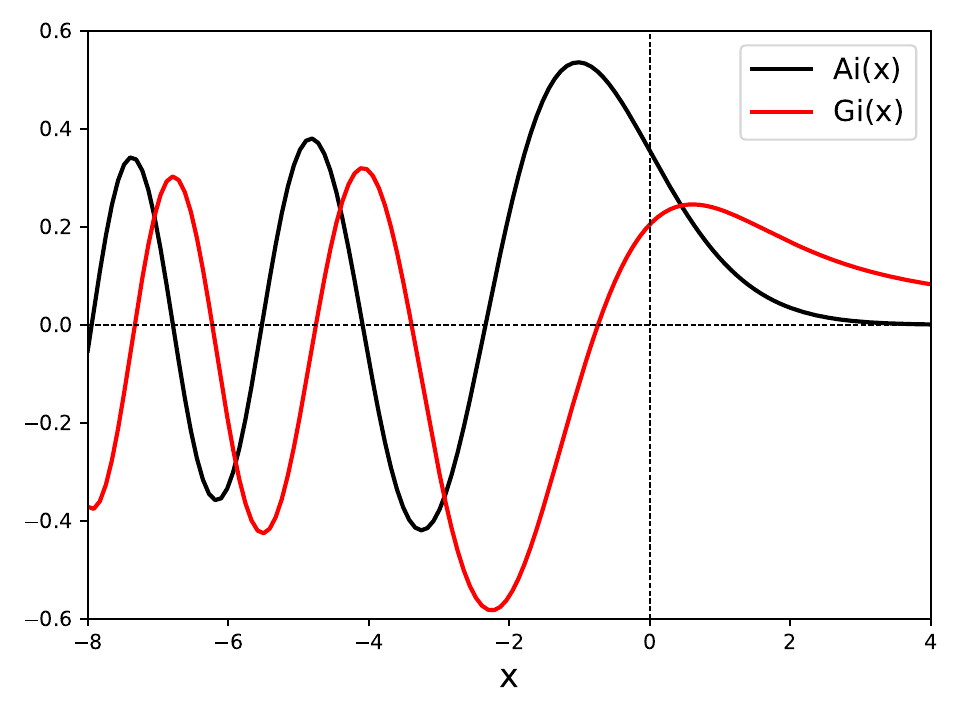}
	\caption{The Airy function $\airyA(x)$ and the Scorer function $\airyG(x)$ plotted against their argument. Both functions behave qualitatively similar, exhibiting exponential decay for $x > 0$ and oscillatory behavior (with relative phase shift) for $x < 0$.}
    \label{fig:airy_scorer}
\end{figure}

\noindent where $\airyA$ denotes the Airy function and $\airyG$ denotes the Scorer function. These functions are plotted in \Fig{fig:airy_scorer} for reference.

One therefore obtains
\begin{equation}
	I(\zeta) = 
	\pi
	\left[
		\airyA\left( \zeta \right)
		+ i \airyG\left( \zeta \right)
	\right]
    .
\end{equation}

\noindent This implies that the incoming and outgoing wavefields can be expressed as
\begin{align}
    \psi_\text{in, out}(\Vect{X}) &=
    \pi \int \dd \Vect{K}_x
    \, \fourier{\psi}_0(\Vect{K}_x) e^{i \Vect{K}_x \cdot \Vect{X}}
    \left[
		\airyA\left( |\Vect{K}_x|^2 - L \right)
		\pm i \airyG\left( |\Vect{K}_x|^2 - L \right)
	\right]
    ,
    \label{eq:psiIN_OUT}
\end{align}

\noindent where the top $(+)$ sign corresponds to the incoming wavefield and the bottom $(-)$ sign corresponds to the outgoing wavefield. Again, we emphasize that these are exact relationships.

\subsection{Fourier Inversion to obtain general solution}

Let us now obtain the desired boundary-value solution (in coordinate space) using \Eq{eq:psiIN_OUT}. To do so, let us perform an FT with respect to only the transverse coordinates $\Vect{X}$. Analogous to \Eq{eq:FT}, this transverse FT takes the form: 
\begin{subequations}
    \begin{align}
        \label{eq:transvFT}
        \widehat{\psi}_\text{in}(\Vect{K}_x)
        &= 
        \int \frac{\dd \Vect{X}}{(2\pi)^N } 
        \, \psi_\text{in}(\Vect{X}) e^{- i \Vect{K}_x \cdot \Vect{X}}
        , \\
        \label{eq:inFIELDft}
        \psi_\text{in}(\Vect{X})
        &= 
        \int \dd \Vect{K}_x
        \, \widehat{\psi}_\text{in}(\Vect{K}_x) e^{i \Vect{K}_x \cdot \Vect{X}}
        .
    \end{align}
\end{subequations}

\noindent Then, we can clearly identify from \Eq{eq:psiIN_OUT} the relationship between the transverse FT and the standard FT images of $\psi$:
\begin{align}
    \widehat{\psi}_\text{in}(\Vect{K}_x)
    &=
    \pi \fourier{\psi}_0(\Vect{K}_x)
    \left[
		\airyA\left( |\Vect{K}_x|^2 - L \right)
		+ i \airyG\left( |\Vect{K}_x|^2 - L \right)
	\right]
    .
    \label{eq:FTrelation}
\end{align}

\noindent Since the quantity $\airyA\left( |\Vect{K}_x|^2 - L \right) + i \airyG\left( |\Vect{K}_x|^2 - L \right)$ is always nonzero (see \App{app:interlacing}), we can then invert the relationship \eq{eq:FTrelation} to obtain
\begin{equation}
	\fourier{\psi}_0(\Vect{K}_x)
    =
    \frac{1}{\pi}
    \frac{
        \widehat{\psi}_\text{in}(\Vect{K}_x)
    }{
        \airyA\left( |\Vect{K}_x|^2 - L \right)
		+ i \airyG\left( |\Vect{K}_x|^2 - L \right)
    }
    .
\end{equation}

The general solution can then be written as
\begin{align}
	\psi(\Vect{X},Z)
	&= \int \dd \Vect{K}_x \, 
    \frac{
        2 \airyA\left( |\Vect{K}_x|^2 + Z - L \right)
        \widehat{\psi}_\text{in}(\Vect{K}_x)
    }{
        \airyA\left( |\Vect{K}_x|^2 - L \right)
		+ i \airyG\left( |\Vect{K}_x|^2 - L \right)
    }
    e^{i \Vect{K}_x \cdot \Vect{X}}
    ,
    \label{eq:solMATCHED}
\end{align}

\noindent where $\widehat{\psi}_\text{in}(\Vect{K}_x)$ is the spectrum of the incoming beam, related to the prescribed boundary value via \Eq{eq:transvFT}. In essence, we have determined the arbitrary function $\fourier{\psi}_0(\Vect{K}_x)$ in \Eq{eq:solGENERAL} that matches to a prescribed boundary condition $\psi_\text{in}(\Vect{X})$ \textit{exactly}, without appealing to asymptotic approximations~\citep{Orlov80,Maj09,Lopez23} that necessarily restrict the validity of the resulting expressions.

As a sanity check, one can confirm that \Eq{eq:solMATCHED} reproduces the known result when $\psi_\text{in}(\Vect{X})$ is a constant; in this case, $\widehat{\psi}_\text{in}(\Vect{K}_x) \propto \delta(\Vect{K}_x)$ such that subsequent integration gives $\psi(\Vect{X}, Z) \propto \airyA(Z - L)$ as desired. Also, note that the incoming and outgoing components to \Eq{eq:solMATCHED} at $Z = 0$ can be identified by performing the re-arrangement
\begin{align}
    &\frac{
        2 \airyA\left( |\Vect{K}_x|^2 - L \right)
    }{
        \airyA\left( |\Vect{K}_x|^2 - L \right)
		+ i \airyG\left( |\Vect{K}_x|^2 - L \right)
    }
    =
    1 + 
    \frac{
        \airyA\left( |\Vect{K}_x|^2 - L \right)
		- i \airyG\left( |\Vect{K}_x|^2 - L \right)
    }{
        \airyA\left( |\Vect{K}_x|^2 - L \right)
		+ i \airyG\left( |\Vect{K}_x|^2 - L \right)
    }
    .
\end{align}

\noindent The incoming or outgoing component corresponds to the subsequent integration of the first or second factor, respectively. In particular, one recovers \Eq{eq:inFIELDft} exactly.

\section{Special case: Incident Gaussian beam in two dimensions}
\label{sec:gauss}

Let us now consider the case when the incoming wavefield is a Gaussian beam, which is of practical importance. We will also specialize to only consider $2$-D propagation ($N = 1$), since this will facilitate comparisons with other published formulas in the literature~\citep{Lopez23,Orlov80,Maj09}. Specifically, we take
\begin{equation}
    \psi_\text{in}(X) = 
    E_0 \exp\left(
        i \sqrt{L} \, X \sin \theta
        - i \frac{X^2 \cos^2 \theta}{2 \sqrt{L}\, \compFOCALnorm}
    \right)
    ,
    \label{eq:inBEAM}
\end{equation}

\noindent where $E_0$ is a constant and $\compFOCALnorm$ is the complex beam parameter normalized by the plasma lengthscale $\airyLEN$, with $\Im(\compFOCALnorm) \ge 0$. Note that we have made use of the relationship
\begin{equation}
    \frac{\omega \airySKIN}{c}
    =
    \sqrt{L}
    .
    \label{eq:normRELATE}
\end{equation}

\noindent Note also that one has
\begin{equation}
    - \frac{i}{\compFOCALnorm}
    =
    - i \frac{\Re(\compFOCALnorm)}{|\compFOCALnorm|^2}
    - \frac{ \Im(\compFOCALnorm)}{|\compFOCALnorm|^2}
    .
    \label{eq:compDEF}
\end{equation}

\noindent Hence, one can identify $|\compFOCALnorm|^2/\Re(\compFOCALnorm)$ as the radius of curvature and $\sqrt{2 L^{-3/2} |\compFOCALnorm|^2/ \Im(\compFOCALnorm)}$ as the beam waist (both normalized by $\airyLEN$). Focusing occurs when $\Re(\compFOCALnorm) > 0$. The linear phase term in \Eq{eq:inBEAM} simply rotates the phasefronts according to the angle of incidence~$\theta$ (with $\theta = 0$ being normal incidence), while the additional factor of $\cos^2 \theta$ in the quadratic phase term in \Eq{eq:inBEAM} accounts for the stretching that occurs for oblique incidence. It is also worth mentioning that when $\theta \neq 0$, \Eq{eq:inBEAM} corresponds to a well-collimated beam (long Rayleigh range) such that the variation of $\compFOCALnorm$ along the incident boundary $z = 0$ can be neglected~\citep{Belyaev24}.

\subsection{Exact solution}

The transverse spectrum of the incoming wavefield is computed to be
\begin{equation}
    \widehat{\psi}_\text{in}(K_x)
    = 
    \norm
    \exp\left[
        \frac{i}{2} \sqrt{L} \, \compFOCALnorm 
        \frac{(K_x - \sqrt{L} \sin\theta)^2}{\cos^2 \theta}
    \right]
    ,
    \label{eq:inSPECTRUM}
\end{equation}

\noindent where $\norm \doteq E_0\sqrt{\frac{\sqrt{L}\, \compFOCALnorm}{2\pi i \cos^2 \theta}}$ is the overall constant. Equation \eq{eq:solMATCHED} therefore takes the form
\begin{align}
    \psi(X,Z)
	&=
    \norm
    \int \dd K_x \,
    \frac{
        2 \airyA\left( K_x^2 + Z - L \right)
    }{
        \airyA\left( K_x^2 - L \right)
		+ i \airyG\left( K_x^2 - L \right)
    }
    e^{i K_x X}
    \nonumber\\
    &\hspace{18mm}\times
    \exp\left[
        \frac{i}{2} \sqrt{L} \, \compFOCALnorm 
        \frac{(K_x - \sqrt{L} \sin\theta)^2}{\cos^2 \theta}
    \right]
    .
    \label{eq:beamEXACT}
\end{align}

\noindent Equation \eq{eq:beamEXACT} depends on four free parameters, three that characterize the boundary value of the incident beam and one that characterizes the medium. They are: (i) $L$, the medium lengthscale normalized by $\airySKIN$ defined in \Eq{eq:skinDEF}, related to normalization by the vacuum wavelength via \Eq{eq:normRELATE}; (ii) $\Re(\compFOCALnorm)$, which parameterizes the incident radius of curvature normalized by the medium lengthscale; (iii) $\Im(\compFOCALnorm) \ge 0$, which parameterizes the incident beam waist normalized by the medium lengthscale; and (iv) $\theta \in [0, \pi/2)$, the angle of incidence with respect to the direction of inhomogeneity $Z$.

\subsection{Oblique Injection as rigid translation and focal shift}

Although the injection angle is nominally a free parameter, when the complex beam parameter is purely real, $\Im(\compFOCALnorm) = 0$, then the injection angle $\theta$ can also be removed by a coordinate transformation, up to an overall phase. Indeed, by completing the square, the initial condition \eq{eq:inBEAM} can be rewritten as:
\begin{equation}
    \psi_\text{in}(X)
    =
    \mc{G}
    \left(
        X -L \compFOCALnorm \frac{\tan \theta}{\cos \theta},
        L,
        \frac{\compFOCALnorm}{\cos^2 \theta}
    \right)
    \exp\left( \frac{i}{2} L^{3/2} \compFOCALnorm \tan^2 \theta \right)
    \label{eq:psiINoblique}
\end{equation}

\noindent where $\mc{G}$ is the Gaussian profile of the incoming beam at normal incidence:
\begin{equation}
    \mc{G}(X;L,\compFOCALnorm) = 
    E_0
    \exp\left(
        - i \frac{X^2}{2 \sqrt{L}\, \compFOCALnorm}
    \right)
    .
\end{equation}

\noindent Clearly, when $\Im(\compFOCALnorm) = 0$, \Eq{eq:psiINoblique} corresponds to rigid translation $\Delta$ in the transverse $X$ direction and a transformed focal length $q_{c,\text{eff}}$ given by
\begin{equation}
    \Delta = L \compFOCALnorm \frac{\tan \theta}{\cos \theta}
    , \quad
    q_{c,\text{eff}} = \frac{\compFOCALnorm}{\cos^2 \theta}
    .
    \label{eq:obliqueMAPPING}
\end{equation}

\noindent In particular, it was shown in \citet{Lopez23} that critical focusing occurs at normal incidence when $\compFOCALnorm = 2$. Hence, for oblique incidence the critical focusing occurs when
\begin{equation}
    q_{c,crit} = 2\cos^2 \theta
    .
    \label{eq:critFOC}
\end{equation}

\noindent Note that these expressions \eq{eq:obliqueMAPPING} and \eq{eq:critFOC} do not contain the additional Goos-Hanchen and focal shifts contained in the analogous formula from \citet{Lopez23}, since these phenomena are only present with a finite beam waist~\citep{McGuirk77}. Hence, one should use these expressions rather than those of \citet{Lopez23} when the beam waist is sufficiently large.

\subsection{Example: Softened critical focusing with finite beam waist}

As discussed in \citet{Lopez23EPS}, the peak intensity of a critically focused wave [\Eq{eq:critFOC} being satisfied] can exceed the standard Airy intensity peak by orders of magnitude%
\footnote{The ratio of the two intensities formally diverges in the short-wavelength asymptotic limit.}. %
However, this only occurs when the beam waist is formally infinite. At normal incidence, the characteristic `detuning width' to still achieve critical focusing is $\Delta_q = 1/\sqrt{L}$~\citep{Lopez23}; if this is entirely accounted for by a finite beam waist [$\Im(\compFOCALnorm) \neq 0$], then \Eq{eq:obliqueMAPPING} implies that for critical focusing to occur, one must have
\begin{subequations}
    \begin{equation}
        \Im(\compFOCALnorm) \lesssim \frac{\cos^2 \theta}{\sqrt{L}}
        ,
        \label{eq:qBOUND}
    \end{equation}

    \noindent or equivalently in terms of the beam waist $W$ [defined following \Eq{eq:compDEF}]:
    \begin{equation}
        \frac{W}{\airySKIN} \gtrsim \sqrt{L} \, \cos \theta
        .
        \label{eq:wBOUND}
    \end{equation}
\end{subequations}

\noindent To obtain \Eq{eq:wBOUND}, one must assume that $q_{c,crit} \gg \Im(\compFOCALnorm)$ given by \Eq{eq:qBOUND}, which is equivalent to the condition that $L \gg 1$.

\begin{figure}
    \centering
    \includegraphics[width=0.24\linewidth,trim={70mm 4mm 8mm 2mm},clip]{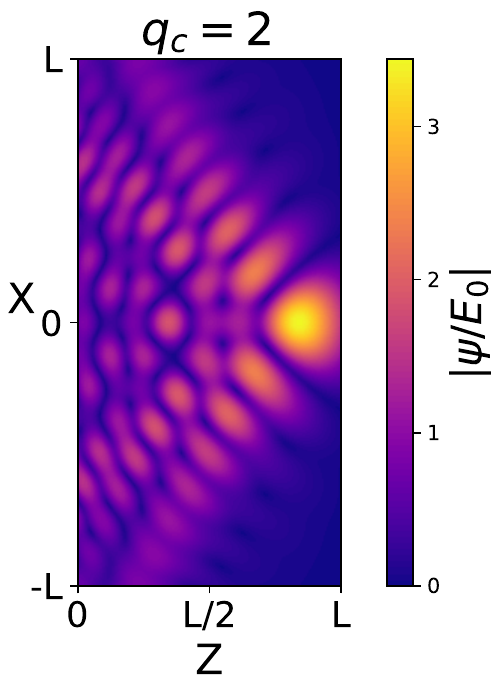}
    \includegraphics[width=0.24\linewidth,trim={70mm 4mm 8mm 2mm},clip]{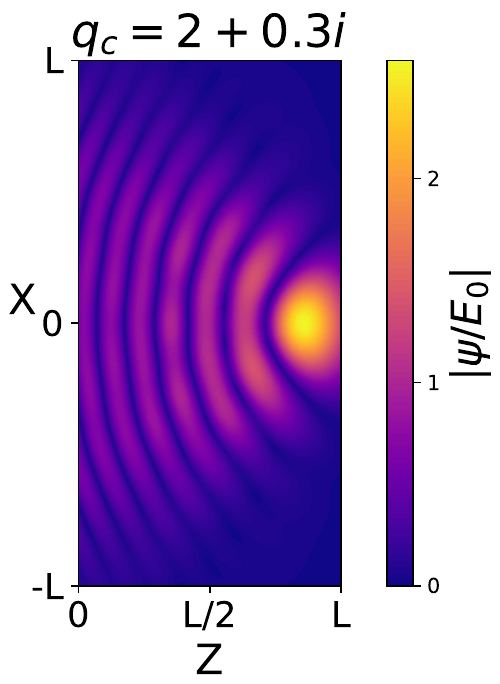}
    \includegraphics[width=0.24\linewidth,trim={68mm 4mm 8mm 2mm},clip]{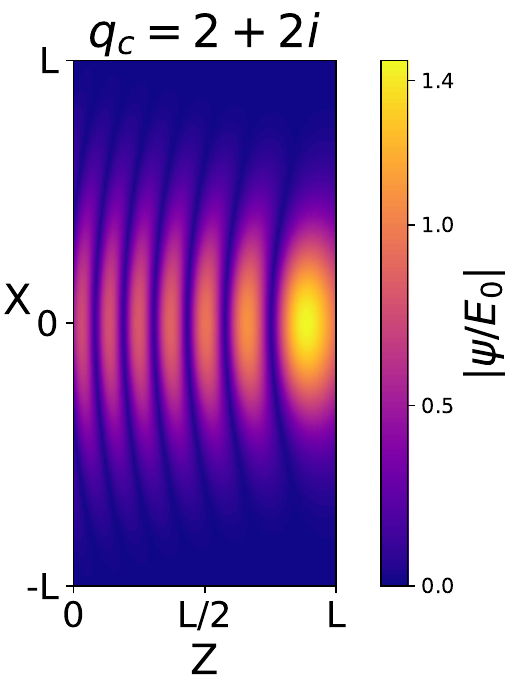}
    \includegraphics[width=0.24\linewidth,trim={66mm 4mm 8mm 2mm},clip]{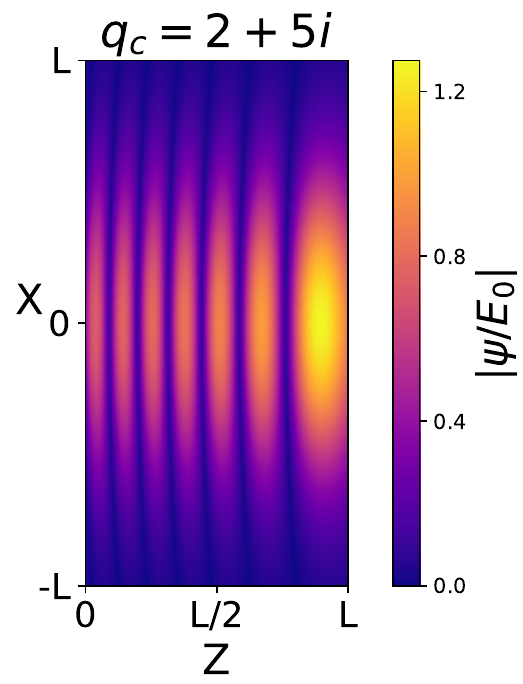}
    \caption{Morphology of the hyperbolic umbilic caustic that occurs at critical focusing \eq{eq:critFOC} as $\Im(\compFOCALnorm)$ is increased from zero at normal incidence ($\theta = 0$) with $L = 10$. The plots are obtained from numerically integrating the exact solution \eq{eq:beamEXACT}. From left to right, the figures show the diffraction pattern at critical focusing, at the expected detuning value of $\Im(\compFOCALnorm)$ \eq{eq:qBOUND}, at the $\Im(\compFOCALnorm)$ that minimizes the beam waist, and at $\Im(\compFOCALnorm) \gg \Re(\compFOCALnorm)$. Note that the colorbar axis changes for each plot.}
    \label{fig:normal}
\end{figure}

\begin{figure}
    \centering
    \includegraphics[width=0.24\linewidth,trim={68mm 4mm 8mm 2mm},clip]{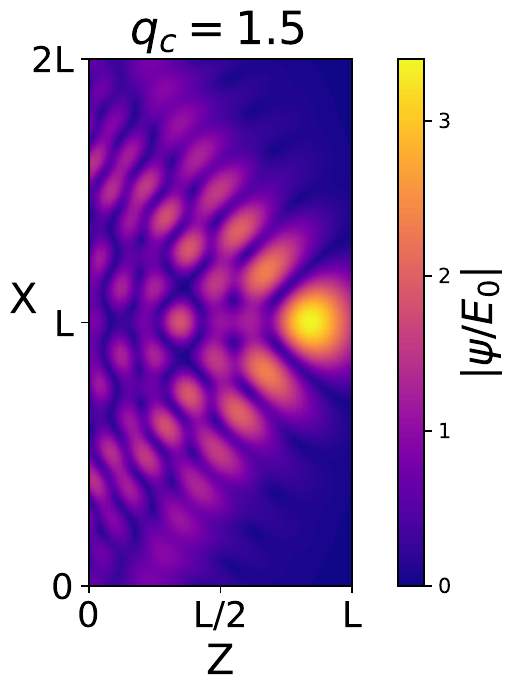}
    \includegraphics[width=0.24\linewidth,trim={66mm 4mm 8mm 2mm},clip]{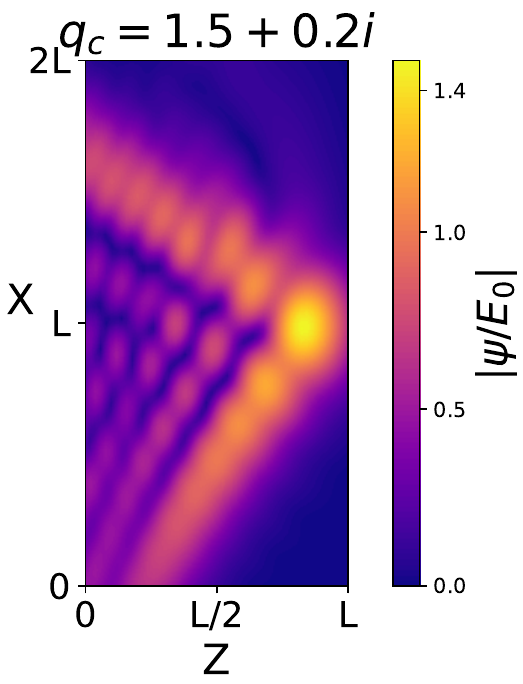}
    \includegraphics[width=0.24\linewidth,trim={64mm 4mm 8mm 2mm},clip]{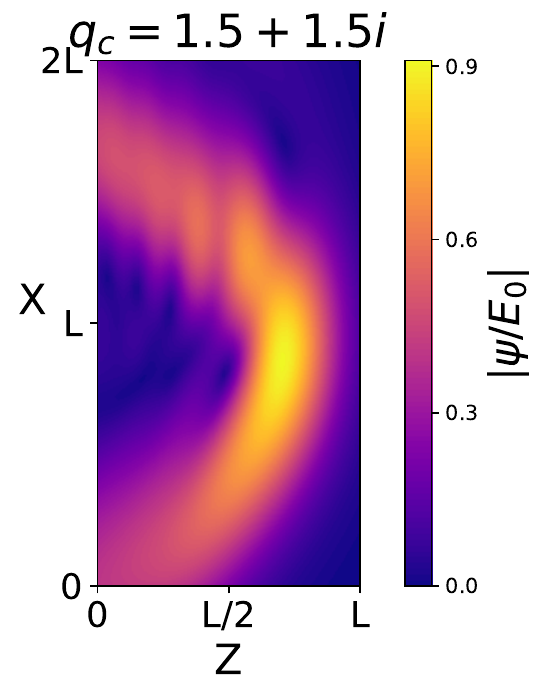}
    \includegraphics[width=0.24\linewidth,trim={64mm 4mm 8mm 2mm},clip]{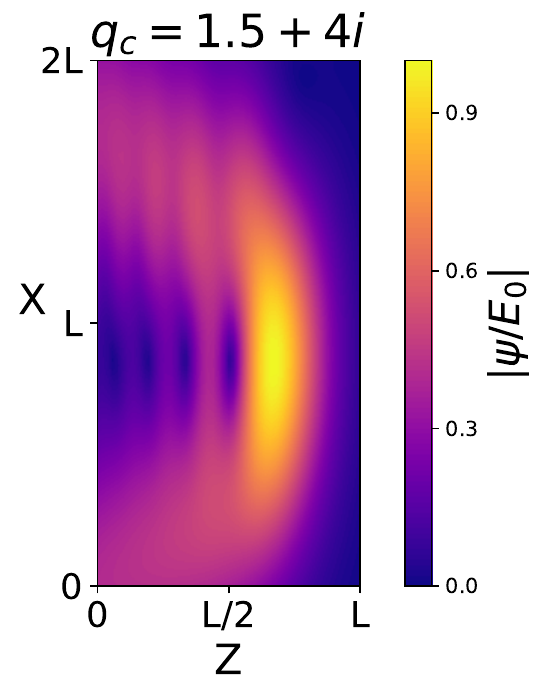}
    \caption{Same as \Fig{fig:normal} but for $\theta = 30^\circ$.}
    \label{fig:oblique}
\end{figure}

\begin{figure}
    \centering
    \includegraphics[width=0.24\linewidth,trim={68mm 4mm 8mm 2mm},clip]{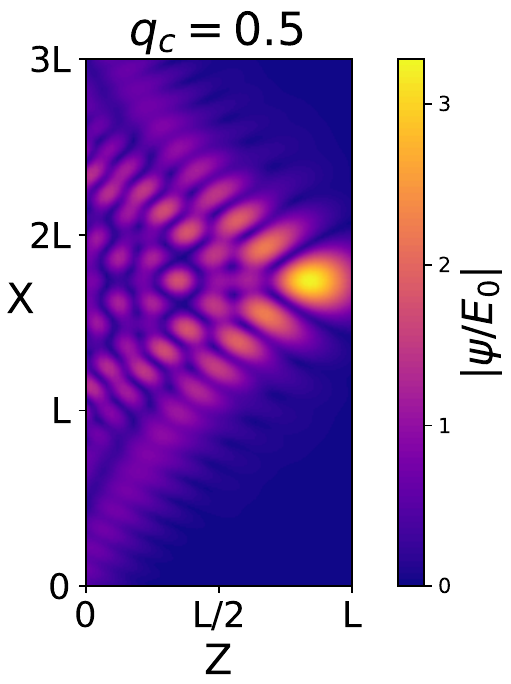}
    \includegraphics[width=0.24\linewidth,trim={66mm 4mm 8mm 2mm},clip]{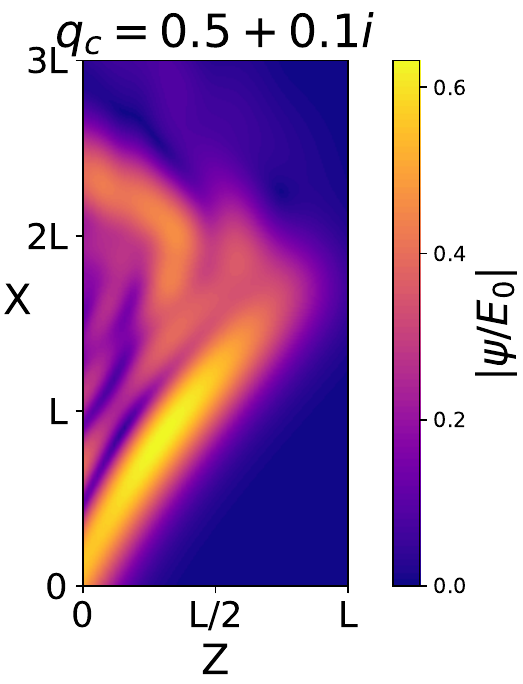}
    \includegraphics[width=0.24\linewidth,trim={66mm 4mm 8mm 2mm},clip]{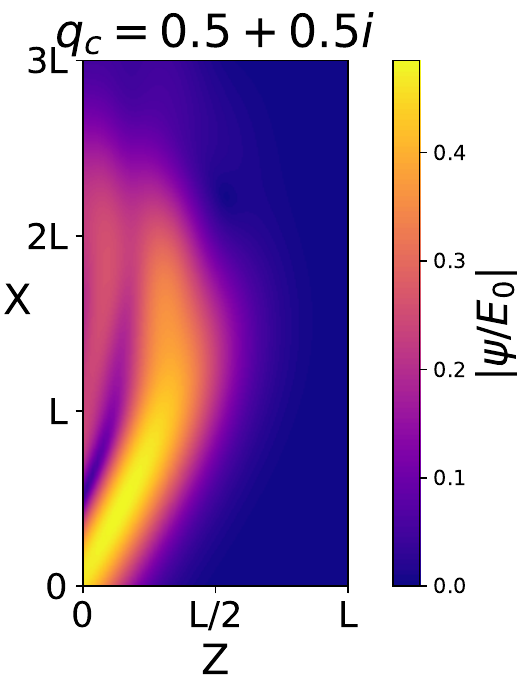}
    \includegraphics[width=0.24\linewidth,trim={64mm 4mm 8mm 2mm},clip]{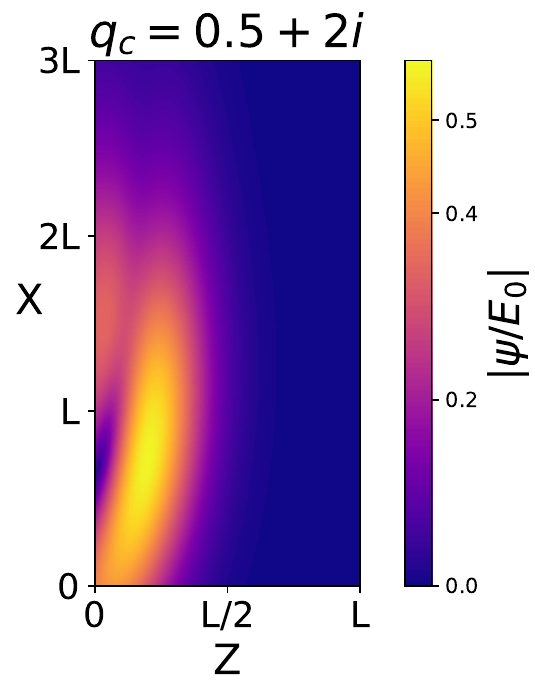}
    \caption{Same as \Fig{fig:normal} but for $\theta = 60^\circ$.}
    \label{fig:shallow}
\end{figure}

Figures \ref{fig:normal} - \ref{fig:shallow} demonstrate how the hyperbolic umbilic diffraction pattern corresponding to critical focusing gets destroyed by a finite beam waist at various values of $\theta$, confirming the prediction of \Eq{eq:wBOUND}. [Note that they also confirm the formulas \eq{eq:obliqueMAPPING} and \eq{eq:critFOC} relating oblique injection with translations and focal shifts for infinitely wide beams.] Also shown are cases when the beam waist is minimized at fixed $\Re(\compFOCALnorm)$ [which occurs for $\Im(\compFOCALnorm) = \Re(\compFOCALnorm)$], and when $\Im(\compFOCALnorm) \gg \Re(\compFOCALnorm)$. By softening the hyperbolic umbilic caustic, the peak intensity decreases with increasing $\Im(\compFOCALnorm)$; this has clear consequences for applications the desire strong focusing near the turning point (such as those discussed in the introduction) and suggests that one should minimize $\Im(\compFOCALnorm)$ as much as possible.

\newpage
\subsection{Comparison with existing asymptotic formulas}

For comparison purposes, let us also list the existing asymptotic formulas provided by \citet{Orlov80}, \citet{Maj09}, and \citet{Lopez23}, which we shall refer to respectively as the O80, the M09, and the L23 formulas. (We reiterate that to the best of our knowledge, only asymptotic solutions to the linear layer problem with prescribed boundary value have appeared in the literature.) These formulas are given for the initial condition \eq{eq:inBEAM} as%
\footnote{Note that we have corrected an error in \citet{Maj09}; their Eq.~(17) is missing an extra factor of $\omega \ell/c$ in the exponent.}
\begin{align}
    \label{eq:OrlovBEAM}
    \psi_\text{O80}(X,Z) &=
    \norm
    \int \dd K_x \,
    \frac{
        2 \airyA\left( K_x^2 + Z - L \right)
    }{
        \airyA\left( K_x^2 - L \right)
		- i \airyA'\left( K_x^2 - L \right)/\sqrt{L - K_x^2 }
    }
    e^{i K_x X}
    \nonumber\\
    &\times
    \exp\left[
        \frac{i}{2} \sqrt{L} \, \compFOCALnorm 
        \frac{(K_x - \sqrt{L} \sin\theta)^2}{\cos^2 \theta}
    \right]
    , \\
    \label{eq:MajBEAM}
    \psi_\text{M09}(X,Z) &=
    \norm 
    \frac{2\pi}{\sqrt{\pi i}} 
    \int_{-\sqrt{L}}^{\sqrt{L}}
    \dd K_x \,
    \left(L - K_x^2 \right)^{1/4}
    \airyA\left( K_x^2 + Z - L \right)
    e^{i K_x X}
    \nonumber\\
    &\times
    \exp\left[
        \frac{i}{2} \sqrt{L} \, \compFOCALnorm 
        \frac{(K_x - \sqrt{L} \sin\theta)^2}{\cos^2 \theta}
        + i \frac{2}{3}
        \left(
            L - K_x^2
        \right)^{3/2}
    \right]
    , \\
    \label{eq:LopezBEAM}
    \psi_\text{L23}(X,Z) &=
    \normLKS 
    \int \dd K_x \,
    \airyA\left(
        K_x^2 + Z - L
    \right)
    e^{i K_x X}
    \nonumber\\
    &\times
    \exp\left[ 
        \frac{i}{2} \sqrt{L} \,
        \frac{
            \compFOCALnorm - 2 \cos 2\theta \cos \theta
        }{
            \cos^2 \theta
        }
        \left(
            K_x
            - \sqrt{L} \,\sin \theta
            \frac{
                \compFOCALnorm + \sin 2\theta \sin \theta
            }{
                \compFOCALnorm - 2 \cos 2\theta \cos \theta
            }
        \right)^2
    \right]
    ,
\end{align}

\noindent where we have introduced
\begin{align}
    \normLKS
    = \norm 
    2\pi 
    \sqrt{
        \frac{\sqrt{L} \, \cos \theta}{\pi i}
    }
    \exp\left(
        \frac{i}{6} L^{3/2} \cos^3 \theta
        \frac{
            4 \compFOCALnorm
            - 7 \cos \theta
            - \cos 3\theta
        }{
            \compFOCALnorm - \cos\theta - \cos 3 \theta
        }
    \right)
    .
\end{align}

\begin{figure}
    \centering
    \includegraphics[width=0.24\linewidth,trim={26mm 6mm 42mm 4mm},clip]{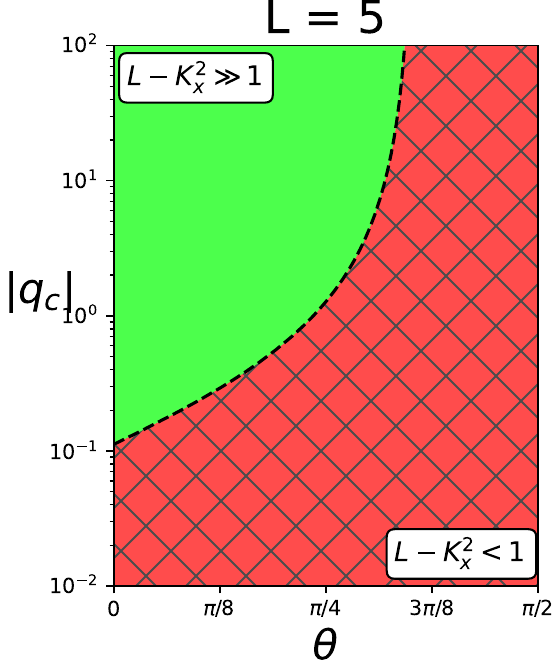}
    \includegraphics[width=0.24\linewidth,trim={26mm 6mm 42mm 4mm},clip]{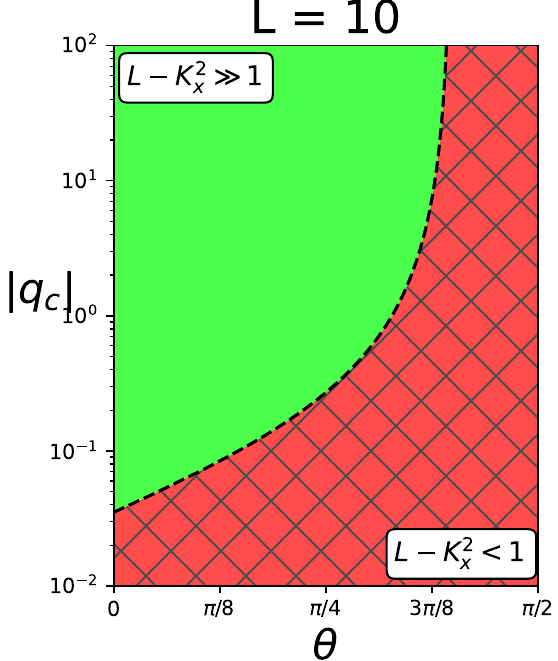}
    \includegraphics[width=0.24\linewidth,trim={26mm 6mm 42mm 4mm},clip]{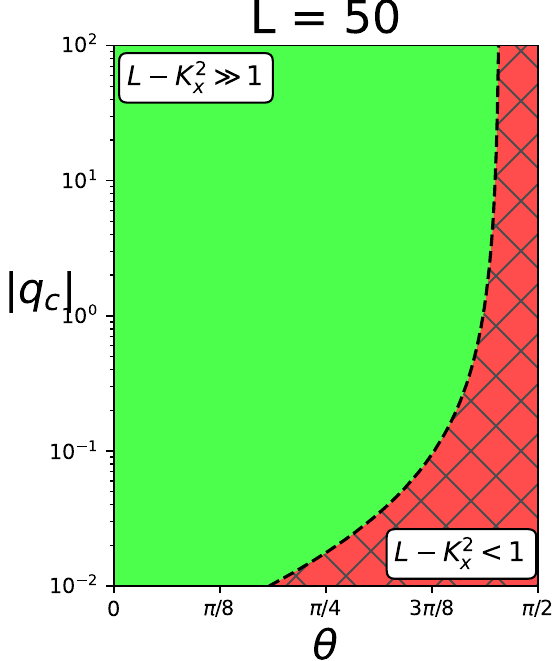}
    \includegraphics[width=0.24\linewidth,trim={26mm 6mm 42mm 4mm},clip]{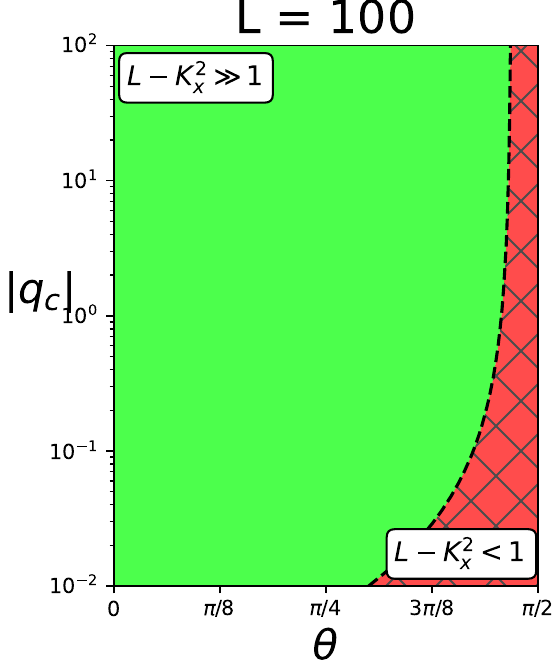}

    \includegraphics[width=0.24\linewidth,trim={26mm 6mm 40mm 2mm},clip]{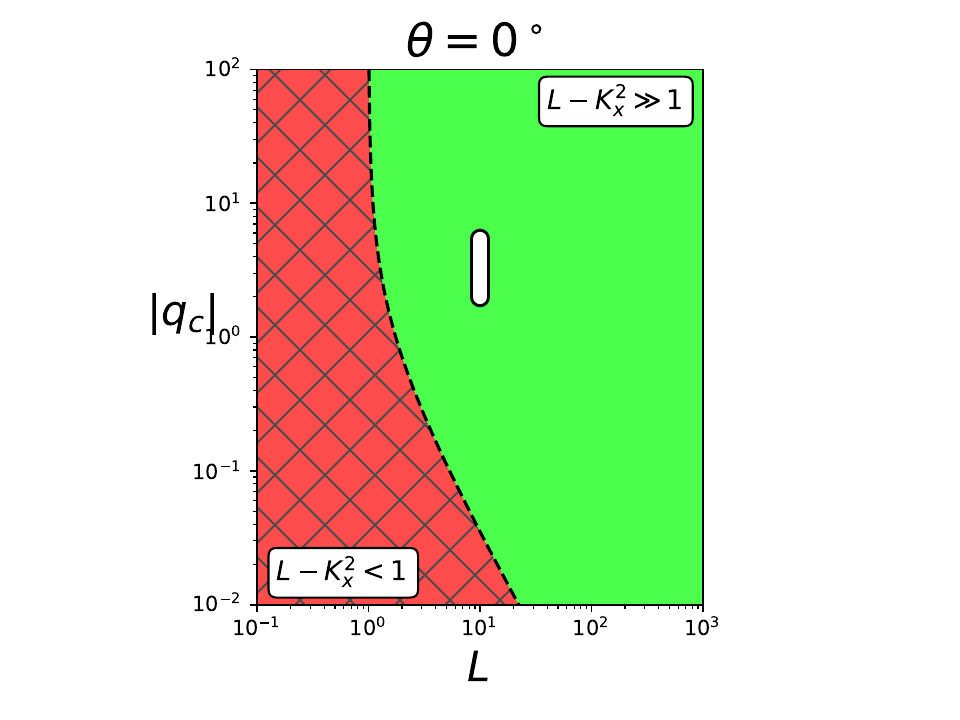}
    \includegraphics[width=0.24\linewidth,trim={26mm 6mm 40mm 2mm},clip]{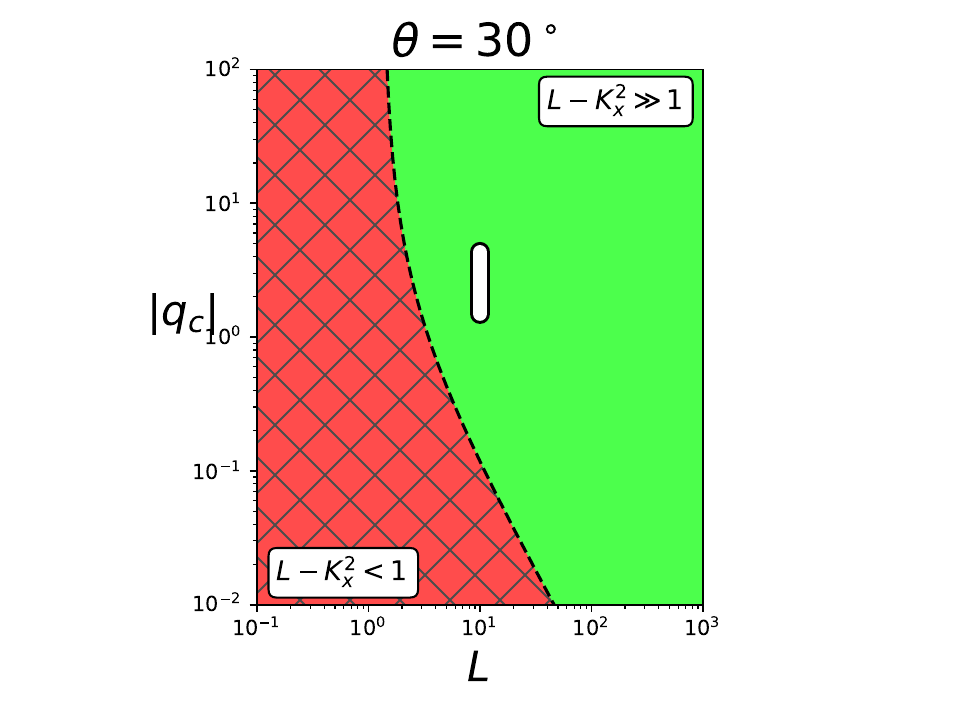}
    \includegraphics[width=0.24\linewidth,trim={26mm 6mm 40mm 2mm},clip]{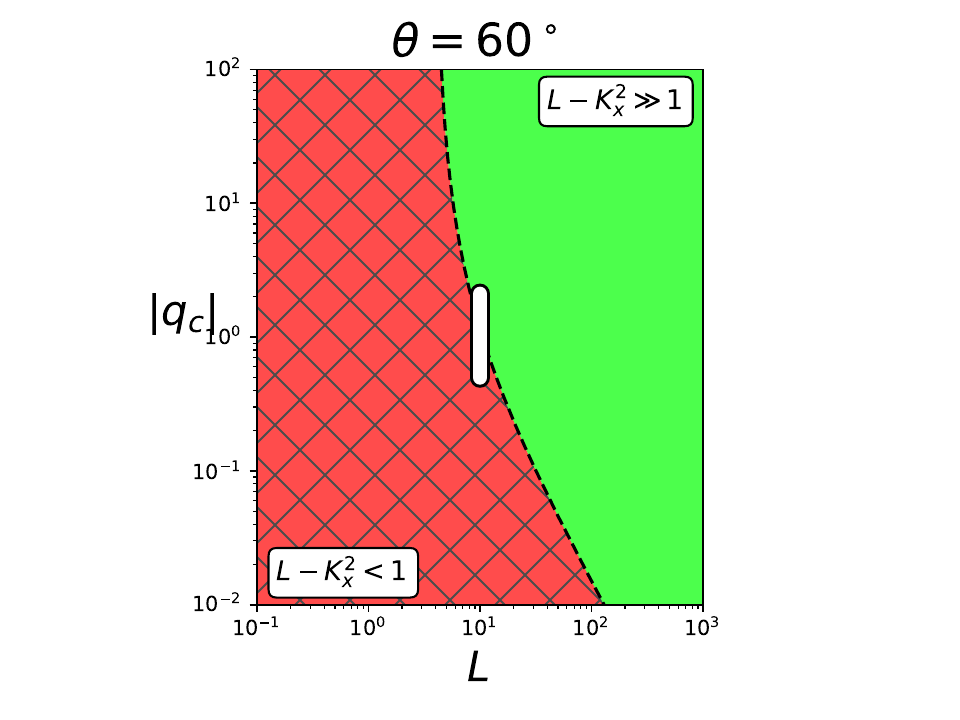}
    \includegraphics[width=0.24\linewidth,trim={26mm 6mm 40mm 2mm},clip]{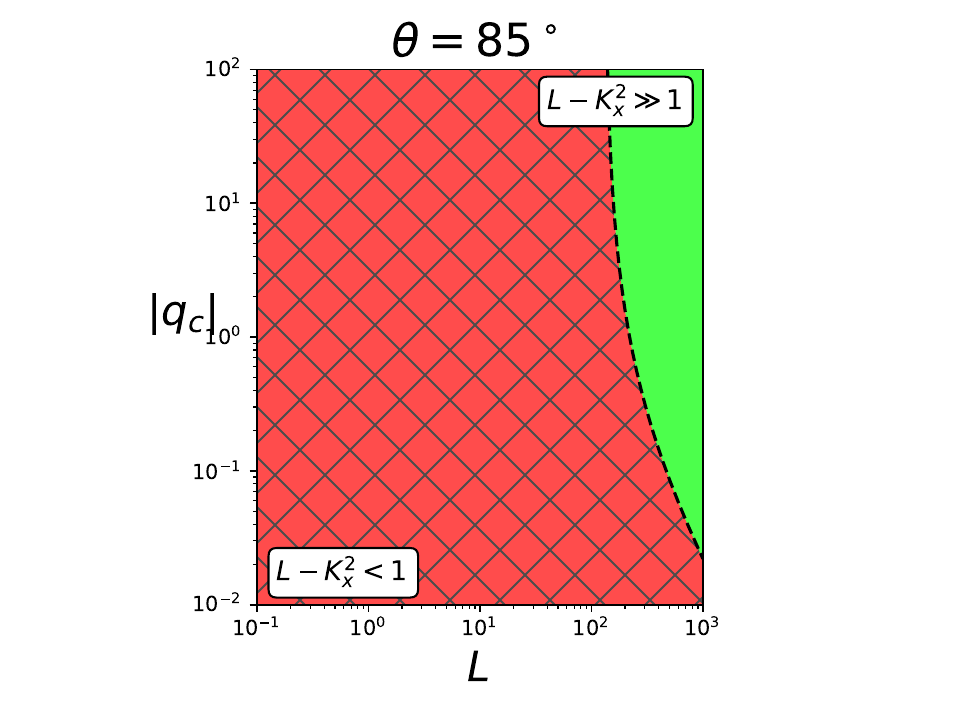}
    \caption{Validity boundary for the asymptotic solutions \eq{eq:OrlovBEAM} - \eq{eq:LopezBEAM} as determined by \Eq{eq:farCOND}, either as a function of $\compFOCALnorm$ and $\theta$ for specified $L$ (top row) or as a function of $\compFOCALnorm$ and $L$ for specified $\theta$ (bottom row). Within the red hatched region, only the exact solution \eq{eq:beamEXACT} remains valid. No valid asymptotic region exists for $L \le 1$ and for $\theta = \pi/2$. The white vertical rectangle in the figures for $\theta = 0^\circ$, $\theta = 30^\circ$, and $\theta = 60^\circ$ corresponds to the range of special cases considered in Figs.~\ref{fig:normal}, \ref{fig:oblique}, and \ref{fig:shallow} respectively.}
    \label{fig:valid}
\end{figure}

As discussed in \App{app:asymptEQUIV}, each of the formulas \eq{eq:OrlovBEAM} - \eq{eq:LopezBEAM} are asymptotically equivalent to the exact solution \eq{eq:beamEXACT} when $L$ is much greater than all wavevectors contained within the incoming spectrum. The region of validity for this condition is plotted in \Fig{fig:valid}. That said, for finite $L - K_x^2$, each of the listed formulas have some shortcomings: \Eq{eq:OrlovBEAM} contains a singularity at $L = K_x^2$ that is clearly unphysical; \Eq{eq:MajBEAM} truncates the integral for $K_x^2 \ge L$ and thereby neglects evanescent contributions to the total field; and \Eq{eq:LopezBEAM} performs a subsidiary Taylor expansion in $K_x^2$ to highlight the caustic structure, with additional loss of accuracy expected as a result. This includes the incorrect predictions for the transverse shift $\Delta$ and the focal shift for a wide beam at oblique incidence discussed in the previous section.

\section{Special case: Speckled plane wave at normal incidence}
\label{sec:speckle}

Let us now consider an incoming field obtained via paraxial propagation from a lens aperture with focal length $\rppLOC$ in 2-D. If the field is evaluated at best focus (\ie following a propagation distance $\rppLOC$), then the (far-field) Fraunhofer diffraction formula gives the solution to be%
\footnote{See, for example, the general discussion in \citet{Lopez22t}, and specifically the cascaded system given by the matrix product of their Eq.~3.137 and Eq.~3.146.}
\begin{equation}
    \psi_\text{in}(X)
    =
    E_0
    \int \dd Y
    \, \psi_0(Y)
    \exp\left(
        - i \frac{\sqrt{L}}{D} X Y
    \right)
    ,
    \label{eq:speckleIN}
\end{equation}

\noindent where $D = d/\airySKIN$ and $\psi_0$ is the field profile that illuminates the lens aperture. Let us choose $\psi_0$ to be a uniformly illuminated RPP array that consists of $\numRPP$ identical elements of width $\width/\numRPP$ (so the total width is $\width$, normalized by $\airySKIN$ as usual):
\begin{equation}
    \psi_0(Y)
    =
    \sum_{\enumRPP=-\frac{\numRPP - 1}{2}}^{\frac{\numRPP - 1}{2}}
    e^{i \thetaRPP}
    \, \text{rect}\left(
        \frac{\numRPP}{\width} Y
        - \enumRPP
    \right)
    ,
\end{equation}

\noindent where $\thetaRPP$ is the corresponding phase shift, set to be either $0$ or $\pi$~\citep{Dixit93}. We take $\numRPP$ to be odd for simplicity. Also, $\text{rect}(x)$ is the unit rectangular function that is nonzero only within the interval $-1/2 \le x \le 1/2$. One then has
\begin{align}
    \widehat{\psi}_\text{in}(K_x)
    =
    E_0
    \frac{D}{\sqrt{L}}
    \sum_{\enumRPP=1}^{\numRPP}
    e^{i \thetaRPP}
    \, \text{rect}\left(
        \frac{\numRPP}{\coupling} K_x
        + \enumRPP
    \right)
    ,
\end{align}

\noindent where we have introduced the coupling coefficient $\coupling$ as
\begin{equation}
    \coupling
    = \frac{\sqrt{L}}{\fnum}
    , \quad
    \fnum \doteq \frac{D}{\width}
    .
    \label{eq:couplingDEF}
\end{equation}

\noindent Note that $\fnum$ is the f-number of the launching aperture. Consequently, \Eq{eq:solMATCHED} reads
\begin{align}
    \psi(X,Z)
	&= 
    E_0
    \frac{D}{\sqrt{L}}
    \sum_{\enumRPP=1}^{\numRPP}
    \exp\left(
        i \thetaRPP
        + i \coupling X \frac{\enumRPP}{M}
    \right)
    \nonumber\\
    &\times
    \int_{- \frac{\coupling}{2 \numRPP}}^{\frac{\coupling}{2 \numRPP}} 
    \dd K_x \, 
    \frac{
        2 \airyA\left[ (K_x + \coupling \frac{\enumRPP }{M})^2 + Z - L \right]
    }{
        \airyA\left[ (K_x + \coupling \frac{\enumRPP }{M})^2 - L \right]
		+ i \airyG\left[ (K_x + \coupling \frac{\enumRPP }{M})^2 - L \right]
    }
    e^{i K_x X}
    .
    \label{eq:speckleEXACT}
\end{align}

Equation \eq{eq:speckleEXACT} is the exact solution for the incoming speckled wavefield given in \Eq{eq:speckleIN}. A complete statistical study of this solution is beyond the scope of the present paper, but one property can be seen immediately. Let us take the number of RPP elements to be sufficiently large, $\numRPP \gg \coupling$, such that the integral in \eq{eq:speckleEXACT} can be approximated by its central value. This yields
\begin{align}
    \psi(X,Z)
	&\approx
    E_0
    \frac{D}{\sqrt{L}}
    \sum_{\enumRPP=1}^{\numRPP}
    \exp\left(
        i \thetaRPP
        + i \coupling X \frac{\enumRPP}{M}
    \right)
    \frac{
        2 \airyA\left[ (\coupling \frac{\enumRPP }{M})^2 + Z - L \right]
    }{
        \airyA\left[ (\coupling \frac{\enumRPP }{M})^2 - L \right]
		+ i \airyG\left[ (\coupling \frac{\enumRPP }{M})^2 - L \right]
    }
    .
\end{align}

\noindent Hence when the coupling is small,
\begin{equation}
    \coupling \ll 1
    ,
\end{equation}

\noindent the transverse speckle pattern and the longitudinal Airy pattern are decoupled from each other. In this limit one needs to simply multiply a given incoming speckle pattern with a single Airy profile along $Z$ to obtain the full solution. 

\begin{figure}
    \centering
    \includegraphics[width=0.32\linewidth,trim={2mm 2mm 10mm 10mm},clip]{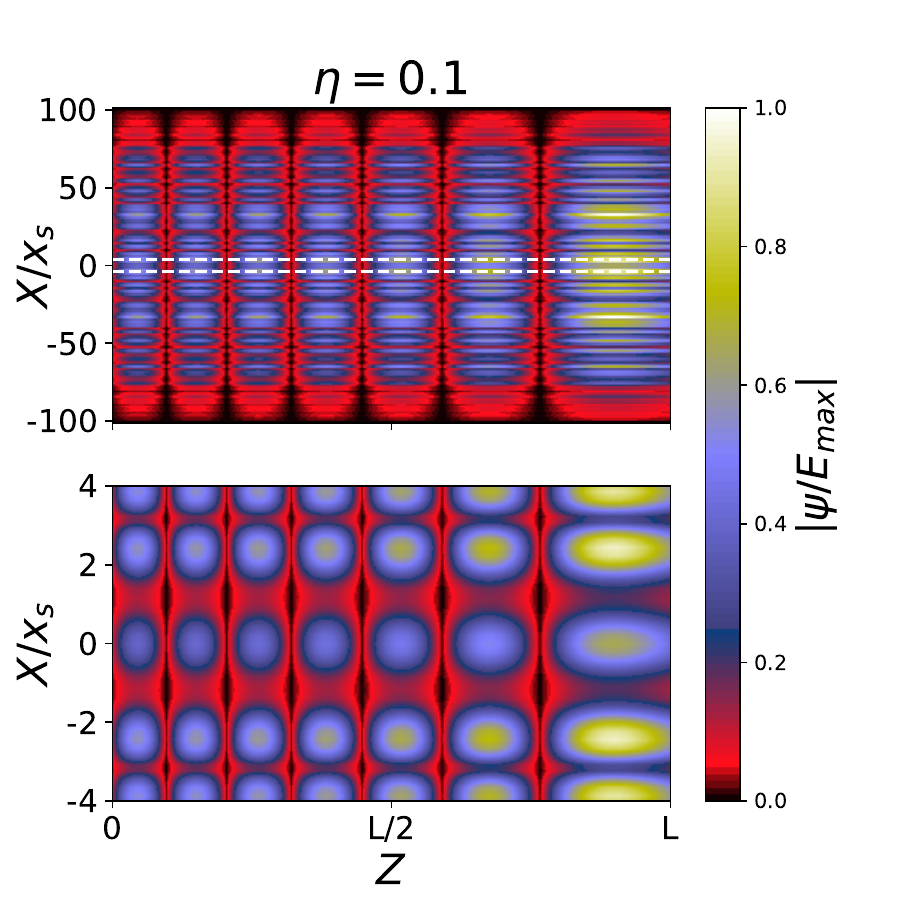}
    \includegraphics[width=0.32\linewidth,trim={2mm 2mm 10mm 10mm},clip]{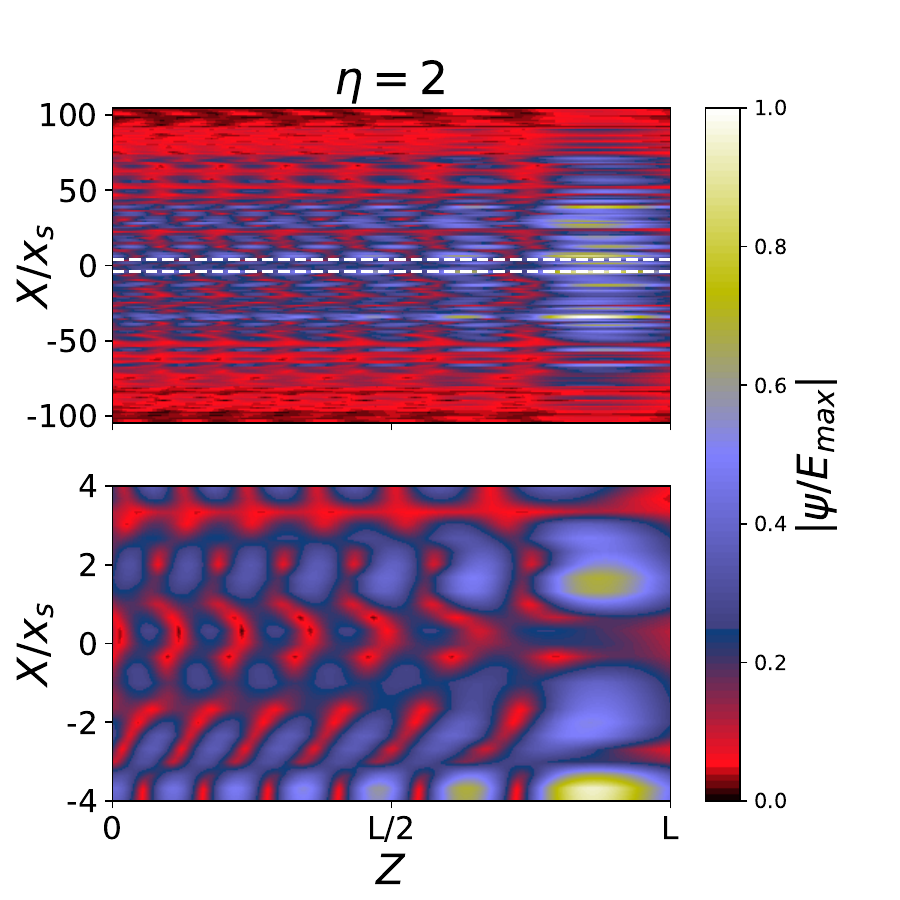}
    \includegraphics[width=0.32\linewidth,trim={2mm 2mm 10mm 10mm},clip]{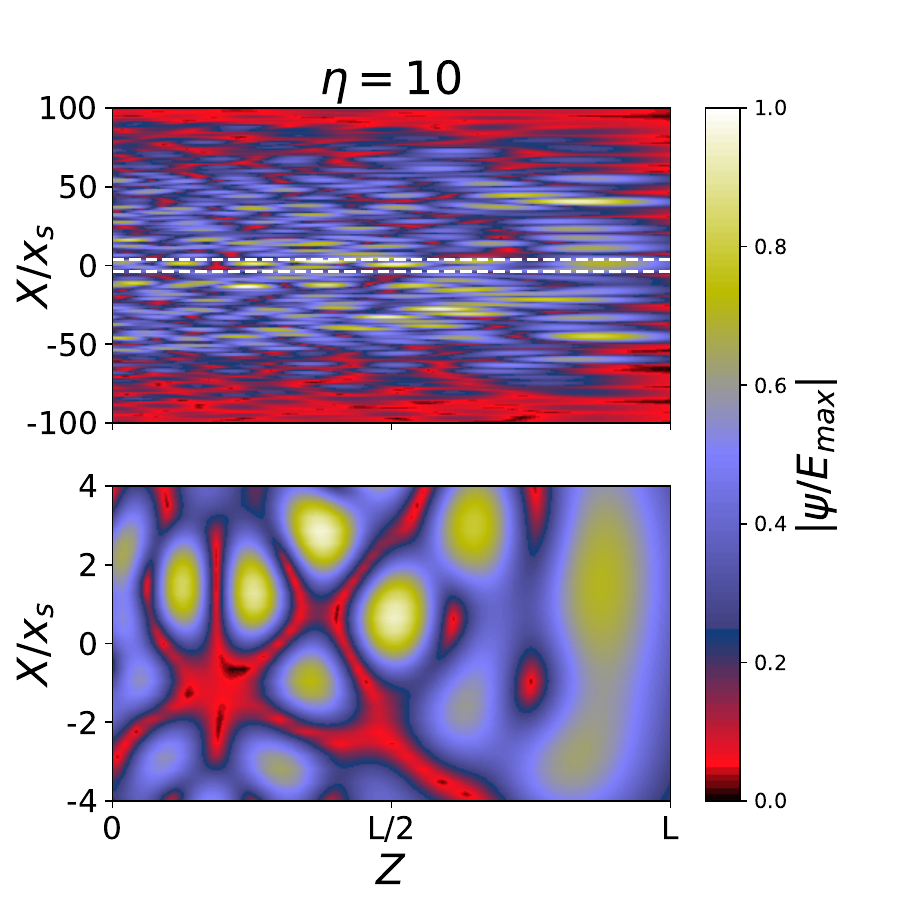}
    \caption{Exact solution \eq{eq:speckleEXACT} for an incoming speckled wavefield at various values of the coupling coefficient $\coupling$ and $L = 10$ and $\numRPP = 100$ RPP elements. The transverse direction is normalized by the nominal speckle width~\citep{Michel23}, which in the normalized coordinates is given by $x_s \sim 2\pi/\coupling$. For each set of figures, the top figure shows the beam profile over the nominal envelope width $M x_s$, while the bottom figure shows the region within the white dashed lines of the top figure.}
    \label{fig:speckleCOMP}
\end{figure}

\begin{figure}
    \centering
    \includegraphics[width=0.32\linewidth,trim = {4mm 4mm 4mm 4mm}, clip]{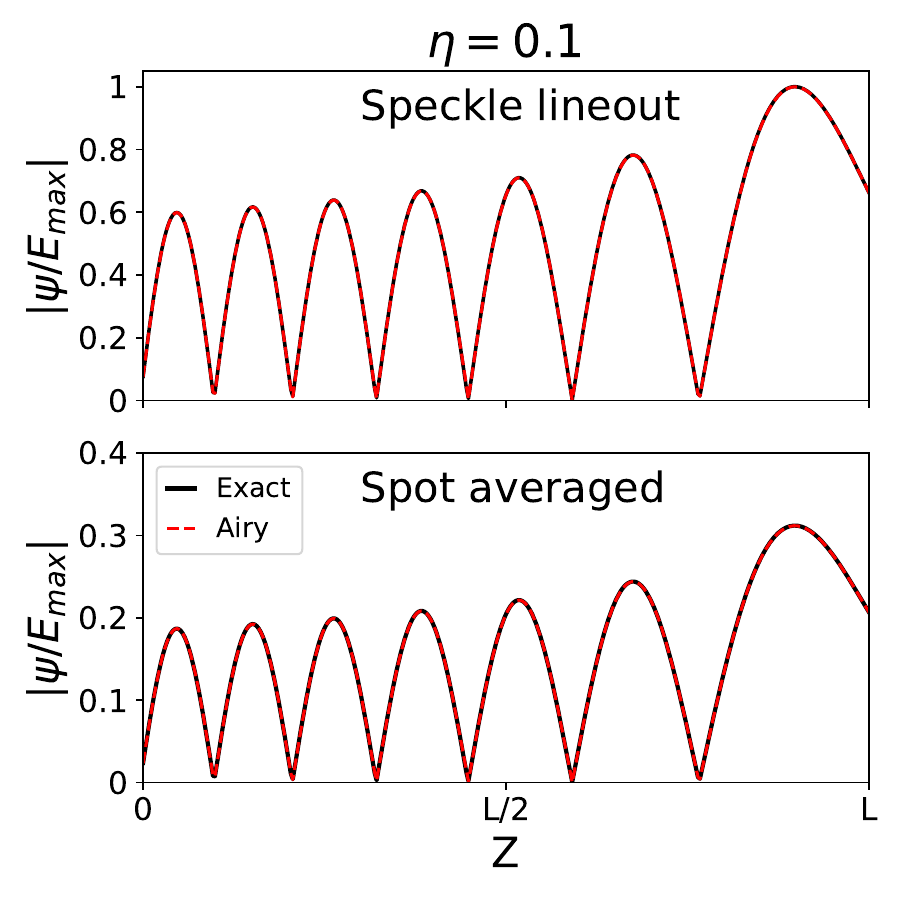}
    \includegraphics[width=0.32\linewidth,trim = {4mm 4mm 4mm 4mm}, clip]{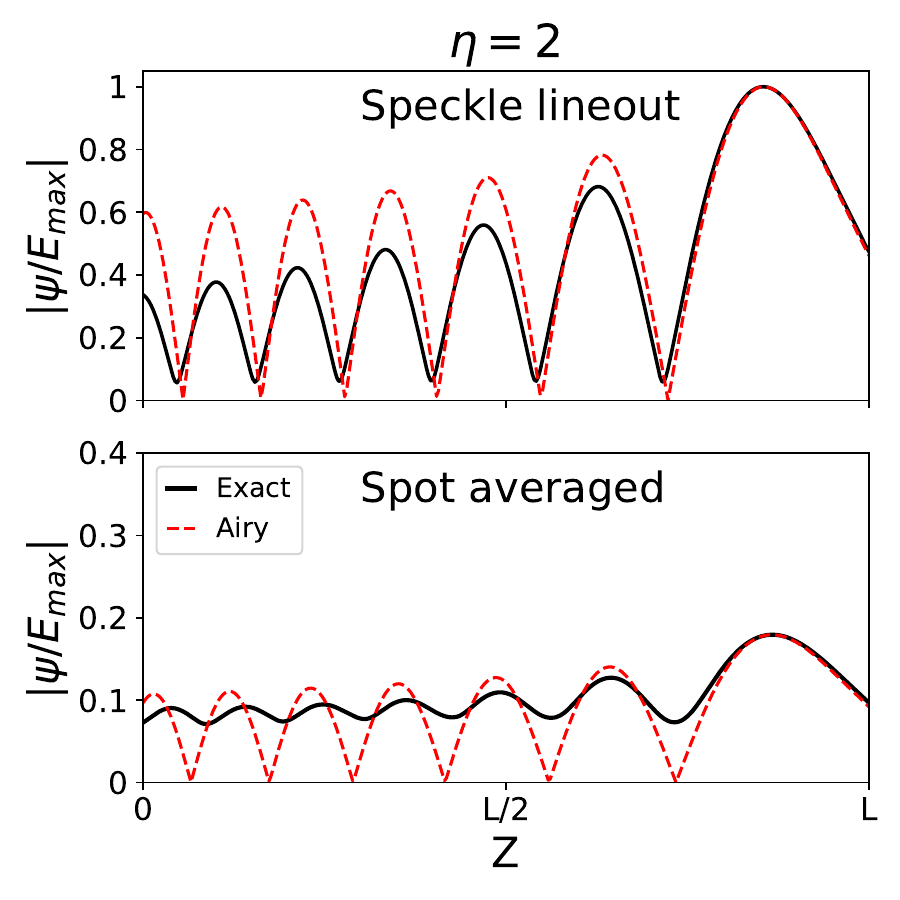}
    \includegraphics[width=0.32\linewidth,trim = {4mm 4mm 4mm 4mm}, clip]{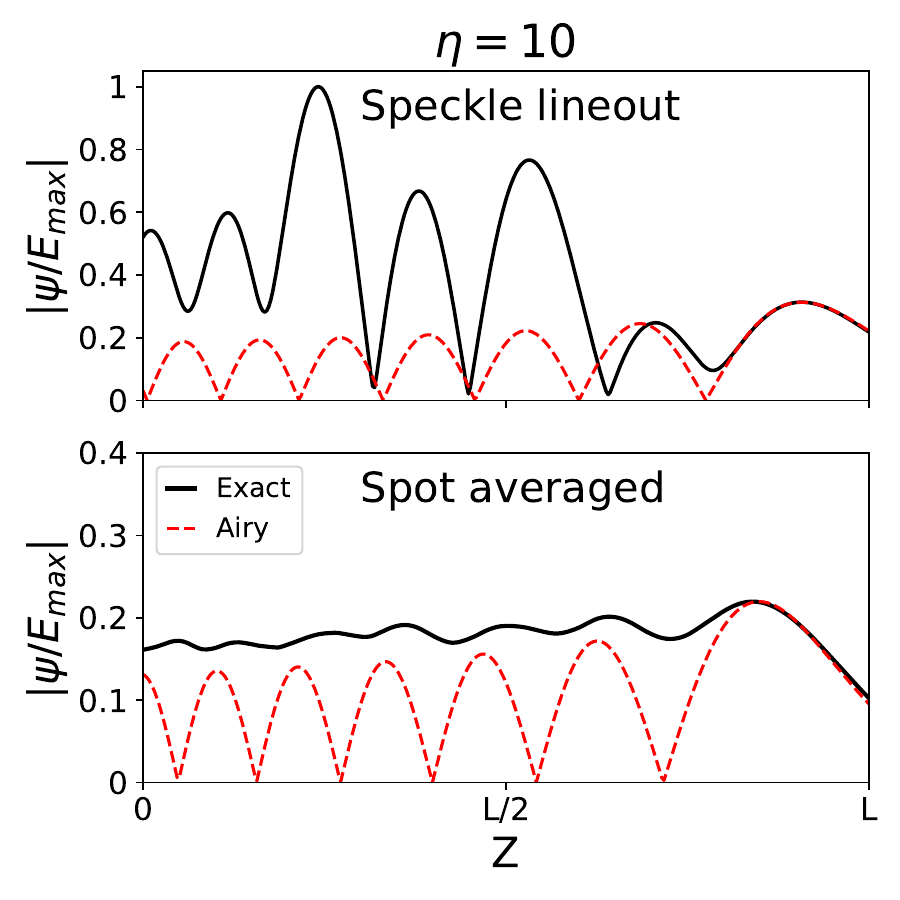}
    \caption{Lineouts along $Z$ (top) of the exact solution presented in \Fig{fig:speckleCOMP} at the location of the brightest speckle, along with the solution averaged over the entire $X$ range (bottom). Also shown are Airy function fits obtained by matching the first peak value and location for each respective plot.}
    \label{fig:speckleLINEOUT}
\end{figure}

As shown by the definition of $\coupling$ in \Eq{eq:couplingDEF}, decoupling occurs when a large f-number beam is launched into a plasma with relatively short lengthscale. Conversely, a small f-number beam launched into a plasma with shallow gradient will exhibit a significantly more complicated longitudinal profile. This is demonstrated in \Fig{fig:speckleCOMP} and \Fig{fig:speckleLINEOUT}, which compare the solution \eq{eq:speckleEXACT} for a larger and smaller coupling coefficient. The decoupled speckles exhibit a regular periodicity along the propagation direction, while the coupled speckles exhibit a considerably more complicated interference pattern. The longitudinal swelling of the coupled speckles also differs significantly from the standard Airy swelling, with the maximum intensity no longer necessarily occurring near the turning point.

\section{Conclusion}
\label{sec:concl}

In this work, the classic problem of an electromagnetic wave propagating in a linearly varying index of refraction (the `linear-layer problem') is revisited. Specifically, we consider what the wavefield throughout the domain is given a prescribed incoming field at the boundary transverse to the direction of inhomogeneity. Previous studies have only obtained asymptotic solutions to this problem setup; here we obtain the exact result by using a spectral matching scheme. The resulting solution \eq{eq:solMATCHED} is expressed as an integral involving the Fourier transform of the prescribed incoming boundary field multiplied by a kernel involving the familiar Airy function and also the related Scorer function. The kernel is never singular and therefore is uniformly valid regardless the relative sizes of the launched wavelength and the medium lengthscale.

An incident Gaussian beam is then studied as a test case, with corresponding exact solution given by \Eq{eq:beamEXACT}. It is shown that when the beam is sufficiently wide (\ie a Gaussian-focused plane wave), oblique angle of incidence results in a simple shift of the focal length and a rigid translation of the diffraction pattern. Explicit expressions are provided by \Eqs{eq:obliqueMAPPING} and \eq{eq:critFOC}, which modify the analogous expressions presented in \citet{Lopez23} by removing the Goos--Hanchen shifts that cannot arise without a finite beam waist. That said, the general prediction of \citet{Lopez23} that the hyperbolic umbilic caustic is structurally stable with respect to angle of incidence still holds true.

It is then demonstrated how the hyperbolic umbilic caustic corresponding to critical focusing of a wide beam gets softened and ultimately disappears as the beam waist is reduced. This softening of the caustic is accompanied by a reduction in the peak intensity obtained by the Gaussian beam at the turning point, which suggests that for applications reliant on the intensity of a wavefield near a turning point, one should make the imaginary part of the complex beam parameter for the launched beam as small as possible. More quantitatively, the hyperbolic umbilic caustic deteriorates once the beam waist $W$ becomes smaller than $W/\airySKIN \lesssim \sqrt{L}$, where $\airySKIN$ is the Airy skin depth~\eq{eq:skinDEF} and $L$ is the medium lengthscale normalized by $\airySKIN$. This means that beam-tracing solutions~\citep{Maj09,Maj10} are fundamentally incapable of describing the hyperbolic umbilic caustic, since the validity of beam-tracing requires $W/\airySKIN \sim \sqrt[4]{L}$ (with $L$ also being large). Advanced ray-tracing methods~\citep{Lopez20,Lopez21a,Lopez22,Lopez24b,Hojlund24} may be able to describe it, however, as they have no such restriction on the beam waist. Conversely, this analysis suggests that any anomalous focusing observed in beam-tracing solutions is not due to the hyperbolic umbilic caustic, but instead due to the relatively simpler Airy (fold) caustic.

Finally, an incident speckled wavefield is also studied. It had been observed previously in numerical parameter scans that large f-number beams in steep gradients experience decoupled speckle and Airy behavior~\citep{Lopez21DPP}, but no concise coupling parameter had been identified. Here we derive the coupling parameter to be $\coupling = \sqrt{L}/\fnum$, and show that speckles only influence the longitudinal profile of the total wavefield when $\coupling \gtrsim 1$, or equivalently, when $\ell \gtrsim \lambda \fnum^3/2\pi$. This suggests that a reduced model of laser beams near turning points is obtained in the low-coupling regime by simply multiplying the transverse speckle pattern with the Airy swelling factor. In the strong-coupling regime, however, the interference patterns are seen to be considerably more complicated, and more work remains to develop reduced models in this limit. To put this finding in context, in terms of the NIF laser~\citep{Spaeth16} the transition to strong coupling occurs for a $351$~nm f/22 beam when the plasma lengthscale exceeds $3.7$~mm, or when it exceeds $0.2$~mm for an f/8 quad beam.

\section*{Acknowledgements}

The author thanks Dr. Juan Ruiz Ruiz of Oxford University for useful conversations. This work is partly supported by STFC (grant number ST/W000903/1).

\appendix

\section{Interlacing of Airy and Scorer zeros}
\label{app:interlacing}

It is important to note
\begin{equation}
    \airyA(x) \pm i \airyG(x) \neq 0
\end{equation}

\noindent because the zeros of $\airyA$ and $\airyG$ never overlap. Since this fact of $\airyG$ is difficult to find in the literature%
\footnote{For example, \citet{Gil03} only compares the zeros of $\airyG$ and $\airyB$, not $\airyG$ and $\airyA$ as we require here.},
we shall present a simple proof here for completeness.

By definition, one has~\citep{Olver10a}
\begin{equation}
	\airyG(x)
	= \airyB(x) \int_x^\infty \airyA(z) \dd z
	+ \airyA(x) \int_0^x \airyB(z) \dd z
    .
\end{equation}

\noindent If one evaluates $\airyG$ at a zero of $\airyA$, denoted $x_0$ (and one notes that $x_0 < 0$), one obtains
\begin{equation}
	\airyG(x_0)
	= \airyB(x_0) \int_{x_0}^\infty \airyA(z) \dd z
 .
\end{equation}

\begin{figure}
	\centering\includegraphics[width=0.6\linewidth]{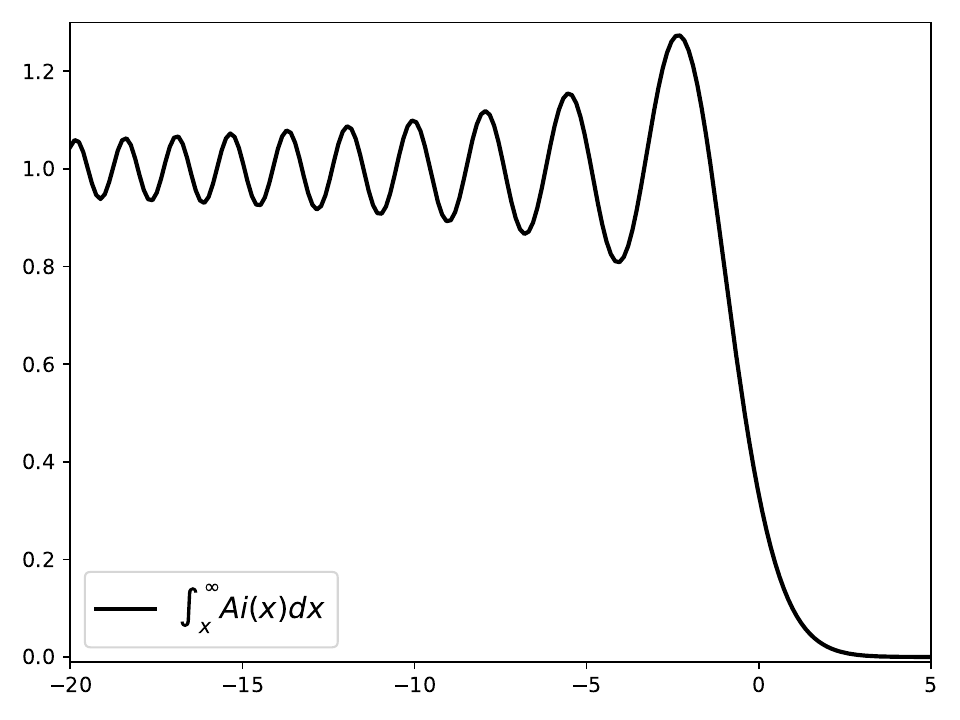}
	\caption{Evaluating the integral $\int_{x}^\infty \airyA(z) \dd z$ for different values of the lower limit $x$. Note that since $\airyA(\zeta) > 0$ for $\zeta > 0$, the integral is manifestly positive for $x > 0$ and asymptotes to zero as $x \to +\infty$.}
    \label{fig:airy_interlace}
\end{figure}

\noindent Since the zeros of $\airyA$ and $\airyB$ are interlaced, one has $\airyB(x_0) \neq 0$. Also, numerical investigation (\Fig{fig:airy_interlace}) shows that
\begin{equation}
	\int_{x}^\infty \airyA(z) \dd z > 0
\end{equation}

\noindent for all values of $x$. Hence $\airyG(x_0) \neq 0$. QED.

\section{Asymptotic equivalence of the exact solution with other published expressions}
\label{app:asymptEQUIV}

Here we show that the exact solution \eq{eq:beamEXACT} is asymptotically equivalent to the other expressions listed in \Eqs{eq:OrlovBEAM} - \eq{eq:LopezBEAM} in the large $L \to \infty$ limit. Specifically, we take $L$ to be much larger than all $K_x^2$ that appear in the incoming spectrum, requiring
\begin{equation}
    L 
    \left[ 
        1
        - \left(
            \sin \theta
            + \frac{\cos \theta}{L^{3/4} \sqrt{|\compFOCALnorm|}}
        \right)^2
    \right]
    \gg 1
    .
    \label{eq:farCOND}
\end{equation}

\noindent Note that to derive \Eq{eq:farCOND}, we have estimated the mean wavevector to be $\sqrt{L} \sin \theta$ and the spectral width to be $\cos \theta / |\compFOCALnorm|^{1/2} L^{1/4}$, per \Eq{eq:inSPECTRUM}. When \Eq{eq:farCOND} is satisfied, we can take $\zeta \doteq L - K_x^2$ to be sufficiently large, viz. $\zeta \gg 1$, such that the following asymptotic approximations hold:
\begin{subequations}
    \label{eq:asymptotic}
    \begin{align}
        \airyA(-\zeta) 
        &\sim \frac{
            \cos \left( \frac{2}{3} \zeta^{3/2} - \frac{\pi}{4}\right)
        }{\sqrt{\pi} \zeta^{1/4}}
        , \\
        \airyA'(-\zeta) 
        &\sim \zeta^{1/4}
        \frac{
            \sin \left( \frac{2}{3} \zeta^{3/2} - \frac{\pi}{4}\right)
        }{\sqrt{\pi} }
        , \\
        \airyG(-\zeta) 
        &\sim 
        - \frac{
            \sin \left( \frac{2}{3} \zeta^{3/2} - \frac{\pi}{4}\right)
        }{\sqrt{\pi} \zeta^{1/4}}
        .
    \end{align}
\end{subequations}

\subsection{Asymptotic equivalence with O80 formula}

The exact solution \eq{eq:beamEXACT} and the O80 formula \eq{eq:OrlovBEAM} differ only via the denominator in the integrand; \Eq{eq:beamEXACT} has $\airyA(-\zeta) + i \airyG(-\zeta)$ while \Eq{eq:OrlovBEAM} has $\airyA(-\zeta) - i \airyA'(-\zeta)/\sqrt{\zeta}$. When $\zeta \gg 1$, the asymptotic relations \eq{eq:asymptotic} then give
\begin{subequations}
    \begin{align}
        \label{eq:denomEXPAND}
        \airyA(-\zeta) + i \airyG(-\zeta)
        &\sim
        \frac{
            \exp \left( - i \frac{2}{3} \zeta^{3/2} + i \frac{\pi}{4}\right)
        }{\sqrt{\pi} \zeta^{1/4}}
        , \\
        \airyA(-\zeta) - i \frac{\airyA'(-\zeta)}{\sqrt{\zeta}}
        &\sim
        \frac{
            \exp \left( - i \frac{2}{3} \zeta^{3/2} + i \frac{\pi}{4}\right)
        }{\sqrt{\pi} \zeta^{1/4}}
        .
    \end{align}
\end{subequations}

\noindent The integrands to both formulas have the same asymptotic limit when $\zeta \gg 1$; hence the two formulas should agree when \Eq{eq:farCOND} is satisfied. That said, it is clear that the integrand of \Eq{eq:OrlovBEAM} has a singularity at $\zeta = 0$, so it will not hold uniformly in $\zeta$.

\subsection{Asymptotic equivalence with M09 formula}

Let us again assume that $\zeta \gg 1$. Applying the asymptotic formula \eq{eq:denomEXPAND} to \Eq{eq:beamEXACT} then gives the integrand of \Eq{eq:MajBEAM}. However, \Eq{eq:MajBEAM} also restricts the integration bounds from the entire real line to the interval $K_x \in [-\sqrt{L}, \sqrt{L}]$. This ensures that $\zeta$ remains positive over the integration, but neglects evanescent modes (with long decay lengths) that may contribute to the exact solution. These modes should be absent when \Eq{eq:farCOND} is well-satisfied, but may be present when it is only marginally true.

\subsection{Asymptotic equivalence with L23 formula}

Again, let us take $\zeta \gg 1$ and apply \eq{eq:denomEXPAND} to \Eq{eq:beamEXACT} to obtain the integrand of \Eq{eq:MajBEAM}. We then perform a subsidiary expansion in the limit of small spectral width $\Delta_k \doteq K_x - K_0$ about the mean wavevector $K_0 = \sqrt{L} \, \sin \theta$. This condition reads $L \cos^2 \theta \gg |(\Delta_k + 2 K_0)\Delta_k|$, which is equivalent to \Eq{eq:farCOND} when the same estimate for $\Delta_k$ is used. The lowest-order approximation for the amplitude gives
\begin{equation}
    \zeta^{1/4}
    =
    L^{1/4} \sqrt{\cos \theta}
    + O(\Delta_k)
    ,
    \label{eq:asymptAMP}
\end{equation}

\noindent while a higher-order approximation for the phase gives
\begin{equation}
    \frac{2}{3} \zeta^{3/2}
    =
    \frac{2}{3} L^{3/2} \cos^3 \theta
    - \epsilon L \sin 2 \theta
    - \epsilon^2 \sqrt{L} \, \frac{\cos 2 \theta}{\cos \theta}
    + O(\Delta_k^3)
    .
    \label{eq:asymptPHASE}
\end{equation}

\noindent Following some algebra, it can then be shown that inserting \Eq{eq:asymptAMP} and \eq{eq:asymptPHASE} into \Eq{eq:beamEXACT} recovers \Eq{eq:LopezBEAM}. Therefore, \Eqs{eq:beamEXACT} and \eq{eq:LopezBEAM} are asymptotically equivalent when \Eq{eq:farCOND} is well-satisfied. eq{eq:LopezBEAM} are asymptotically equivalent when \Eq{eq:farCOND} is well-satisfied. 

\bibliography{Biblio.bib}

\begin{thebibliography}{23}%
\makeatletter
\providecommand \@ifxundefined [1]{%
 \@ifx{#1\undefined}
}%
\providecommand \@ifnum [1]{%
 \ifnum #1\expandafter \@firstoftwo
 \else \expandafter \@secondoftwo
 \fi
}%
\providecommand \@ifx [1]{%
 \ifx #1\expandafter \@firstoftwo
 \else \expandafter \@secondoftwo
 \fi
}%
\providecommand \natexlab [1]{#1}%
\providecommand \enquote  [1]{``#1''}%
\providecommand \bibnamefont  [1]{#1}%
\providecommand \bibfnamefont [1]{#1}%
\providecommand \citenamefont [1]{#1}%
\providecommand \href@noop [0]{\@secondoftwo}%
\providecommand \href [0]{\begingroup \@sanitize@url \@href}%
\providecommand \@href[1]{\@@startlink{#1}\@@href}%
\providecommand \@@href[1]{\endgroup#1\@@endlink}%
\providecommand \@sanitize@url [0]{\catcode `\\12\catcode `\$12\catcode `\&12\catcode `\#12\catcode `\^12\catcode `\_12\catcode `\%12\relax}%
\providecommand \@@startlink[1]{}%
\providecommand \@@endlink[0]{}%
\providecommand \url  [0]{\begingroup\@sanitize@url \@url }%
\providecommand \@url [1]{\endgroup\@href {#1}{\urlprefix }}%
\providecommand \urlprefix  [0]{URL }%
\providecommand \Eprint [0]{\href }%
\providecommand \doibase [0]{http://dx.doi.org/}%
\providecommand \selectlanguage [0]{\@gobble}%
\providecommand \bibinfo  [0]{\@secondoftwo}%
\providecommand \bibfield  [0]{\@secondoftwo}%
\providecommand \translation [1]{[#1]}%
\providecommand \BibitemOpen [0]{}%
\providecommand \bibitemStop [0]{}%
\providecommand \bibitemNoStop [0]{.\EOS\space}%
\providecommand \EOS [0]{\spacefactor3000\relax}%
\providecommand \BibitemShut  [1]{\csname bibitem#1\endcsname}%
\let\auto@bib@innerbib\@empty
\bibitem [{\citenamefont {Ginzburg}(1961)}]{Ginzburg61}%
  \BibitemOpen
  \bibfield  {author} {\bibinfo {author} {\bibfnamefont {V.~L.}\ \bibnamefont {Ginzburg}},\ }\href@noop {} {\emph {\bibinfo {title} {Propagation of electromagnetic waves in plasma}}}\ (\bibinfo  {publisher} {New York: Gordon and Breach},\ \bibinfo {year} {1961})\BibitemShut {NoStop}%
\bibitem [{\citenamefont {Ono}\ \emph {et~al.}(2022{\natexlab{a}})\citenamefont {Ono}, \citenamefont {Bertelli}, \citenamefont {Shevchenko}, \citenamefont {Idei},\ and\ \citenamefont {Hanada}}]{Ono22a}%
  \BibitemOpen
  \bibfield  {author} {\bibinfo {author} {\bibfnamefont {M.}~\bibnamefont {Ono}}, \bibinfo {author} {\bibfnamefont {N.}~\bibnamefont {Bertelli}}, \bibinfo {author} {\bibfnamefont {V.}~\bibnamefont {Shevchenko}}, \bibinfo {author} {\bibfnamefont {H.}~\bibnamefont {Idei}}, \ and\ \bibinfo {author} {\bibfnamefont {K.}~\bibnamefont {Hanada}},\ }\href {\doibase 10.1103/PhysRevE.106.L023201} {\bibfield  {journal} {\bibinfo  {journal} {Phys. Rev. E}\ }\textbf {\bibinfo {volume} {106}},\ \bibinfo {pages} {L023201} (\bibinfo {year} {2022}{\natexlab{a}})}\BibitemShut {NoStop}%
\bibitem [{\citenamefont {Ono}\ \emph {et~al.}(2022{\natexlab{b}})\citenamefont {Ono}, \citenamefont {Bertelli},\ and\ \citenamefont {Shevchenko}}]{Ono22b}%
  \BibitemOpen
  \bibfield  {author} {\bibinfo {author} {\bibfnamefont {M.}~\bibnamefont {Ono}}, \bibinfo {author} {\bibfnamefont {N.}~\bibnamefont {Bertelli}}, \ and\ \bibinfo {author} {\bibfnamefont {V.}~\bibnamefont {Shevchenko}},\ }\href {\doibase 10.1088/1741-4326/ac8be2} {\bibfield  {journal} {\bibinfo  {journal} {Nucl. Fusion}\ }\textbf {\bibinfo {volume} {62}},\ \bibinfo {pages} {106035} (\bibinfo {year} {2022}{\natexlab{b}})}\BibitemShut {NoStop}%
\bibitem [{\citenamefont {{Hall-Chen}}\ \emph {et~al.}(2022)\citenamefont {{Hall-Chen}}, \citenamefont {Parra},\ and\ \citenamefont {Hillesheim}}]{HallChen22}%
  \BibitemOpen
  \bibfield  {author} {\bibinfo {author} {\bibfnamefont {V.~H.}\ \bibnamefont {{Hall-Chen}}}, \bibinfo {author} {\bibfnamefont {F.~I.}\ \bibnamefont {Parra}}, \ and\ \bibinfo {author} {\bibfnamefont {J.~C.}\ \bibnamefont {Hillesheim}},\ }\href {\doibase 10.1088/1361-6587/ac57a1} {\bibfield  {journal} {\bibinfo  {journal} {Plasma Phys. Control. Fusion}\ }\textbf {\bibinfo {volume} {64}},\ \bibinfo {pages} {095002} (\bibinfo {year} {2022})}\BibitemShut {NoStop}%
\bibitem [{\citenamefont {Orlov}\ and\ \citenamefont {Tropkin}(1980)}]{Orlov80}%
  \BibitemOpen
  \bibfield  {author} {\bibinfo {author} {\bibfnamefont {Y.~I.}\ \bibnamefont {Orlov}}\ and\ \bibinfo {author} {\bibfnamefont {S.~K.}\ \bibnamefont {Tropkin}},\ }\href {\doibase 10.1007/BF01033467} {\bibfield  {journal} {\bibinfo  {journal} {Radiophys. Quantum Electron.}\ }\textbf {\bibinfo {volume} {23}},\ \bibinfo {pages} {979} (\bibinfo {year} {1980})}\BibitemShut {NoStop}%
\bibitem [{\citenamefont {Maj}\ \emph {et~al.}(2009)\citenamefont {Maj}, \citenamefont {Pereverzev},\ and\ \citenamefont {Poli}}]{Maj09}%
  \BibitemOpen
  \bibfield  {author} {\bibinfo {author} {\bibfnamefont {O.}~\bibnamefont {Maj}}, \bibinfo {author} {\bibfnamefont {G.~V.}\ \bibnamefont {Pereverzev}}, \ and\ \bibinfo {author} {\bibfnamefont {E.}~\bibnamefont {Poli}},\ }\href {\doibase 10.1063/1.3155449} {\bibfield  {journal} {\bibinfo  {journal} {Phys. Plasmas}\ }\textbf {\bibinfo {volume} {16}},\ \bibinfo {pages} {062105} (\bibinfo {year} {2009})}\BibitemShut {NoStop}%
\bibitem [{\citenamefont {Maj}\ \emph {et~al.}(2010)\citenamefont {Maj}, \citenamefont {Balakin},\ and\ \citenamefont {Poli}}]{Maj10}%
  \BibitemOpen
  \bibfield  {author} {\bibinfo {author} {\bibfnamefont {O.}~\bibnamefont {Maj}}, \bibinfo {author} {\bibfnamefont {A.~A.}\ \bibnamefont {Balakin}}, \ and\ \bibinfo {author} {\bibfnamefont {E.}~\bibnamefont {Poli}},\ }\href {\doibase 10.1088/0741-3335/52/8/085006} {\bibfield  {journal} {\bibinfo  {journal} {Plasma Phys. Control. Fusion}\ }\textbf {\bibinfo {volume} {52}},\ \bibinfo {pages} {085006} (\bibinfo {year} {2010})}\BibitemShut {NoStop}%
\bibitem [{\citenamefont {Lopez}\ \emph {et~al.}(2023)\citenamefont {Lopez}, \citenamefont {Kur},\ and\ \citenamefont {Strozzi}}]{Lopez23}%
  \BibitemOpen
  \bibfield  {author} {\bibinfo {author} {\bibfnamefont {N.~A.}\ \bibnamefont {Lopez}}, \bibinfo {author} {\bibfnamefont {E.}~\bibnamefont {Kur}}, \ and\ \bibinfo {author} {\bibfnamefont {D.~J.}\ \bibnamefont {Strozzi}},\ }\href {\doibase 10.1103/PhysRevE.107.055204} {\bibfield  {journal} {\bibinfo  {journal} {Phys. Rev. E}\ }\textbf {\bibinfo {volume} {107}},\ \bibinfo {pages} {055204} (\bibinfo {year} {2023})}\BibitemShut {NoStop}%
\bibitem [{\citenamefont {Michel}(2023)}]{Michel23}%
  \BibitemOpen
  \bibfield  {author} {\bibinfo {author} {\bibfnamefont {P.}~\bibnamefont {Michel}},\ }\href {\doibase 10.1007/978-3-031-23424-8} {\emph {\bibinfo {title} {Introduction to Laser-Plasma Interactions}}}\ (\bibinfo  {publisher} {Cham: Springer},\ \bibinfo {year} {2023})\BibitemShut {NoStop}%
\bibitem [{\citenamefont {Olver}\ \emph {et~al.}(2010)\citenamefont {Olver}, \citenamefont {Lozier}, \citenamefont {Boisvert},\ and\ \citenamefont {Clark}}]{Olver10a}%
  \BibitemOpen
  \bibfield  {author} {\bibinfo {author} {\bibfnamefont {F.~W.~J.}\ \bibnamefont {Olver}}, \bibinfo {author} {\bibfnamefont {D.~W.}\ \bibnamefont {Lozier}}, \bibinfo {author} {\bibfnamefont {R.~F.}\ \bibnamefont {Boisvert}}, \ and\ \bibinfo {author} {\bibfnamefont {C.~W.}\ \bibnamefont {Clark}},\ }\href@noop {} {\emph {\bibinfo {title} {NIST Handbook of Mathematical Functions}}}\ (\bibinfo  {publisher} {Cambridge: Cambridge University Press},\ \bibinfo {year} {2010})\BibitemShut {NoStop}%
\bibitem [{\citenamefont {Belyaev}\ \emph {et~al.}(2024)\citenamefont {Belyaev}, \citenamefont {Banks},\ and\ \citenamefont {Chapman}}]{Belyaev24}%
  \BibitemOpen
  \bibfield  {author} {\bibinfo {author} {\bibfnamefont {M.~A.}\ \bibnamefont {Belyaev}}, \bibinfo {author} {\bibfnamefont {J.}~\bibnamefont {Banks}}, \ and\ \bibinfo {author} {\bibfnamefont {T.}~\bibnamefont {Chapman}},\ }\href {\doibase 10.1063/5.0198523} {\bibfield  {journal} {\bibinfo  {journal} {Phys. Plasmas}\ }\textbf {\bibinfo {volume} {31}},\ \bibinfo {pages} {053901} (\bibinfo {year} {2024})}\BibitemShut {NoStop}%
\bibitem [{\citenamefont {McGuirk}\ and\ \citenamefont {Carniglia}(1977)}]{McGuirk77}%
  \BibitemOpen
  \bibfield  {author} {\bibinfo {author} {\bibfnamefont {M.}~\bibnamefont {McGuirk}}\ and\ \bibinfo {author} {\bibfnamefont {C.~K.}\ \bibnamefont {Carniglia}},\ }\href {\doibase 10.1364/JOSA.67.000103} {\bibfield  {journal} {\bibinfo  {journal} {J. Opt. Soc. Am.}\ }\textbf {\bibinfo {volume} {67}},\ \bibinfo {pages} {103} (\bibinfo {year} {1977})}\BibitemShut {NoStop}%
\bibitem [{\citenamefont {Lopez}(2023)}]{Lopez23EPS}%
  \BibitemOpen
  \bibfield  {author} {\bibinfo {author} {\bibfnamefont {N.~A.}\ \bibnamefont {Lopez}},\ }in\ \href {https://arxiv.org/abs/2309.15108} {\emph {\bibinfo {booktitle} {49th EPS Conference on Plasma Physics, EPS 2023}}}\ (\bibinfo  {publisher} {arXiv:2309.15108},\ \bibinfo {year} {2023})\ p.\ \bibinfo {pages} {P2.011}\BibitemShut {NoStop}%
\bibitem [{\citenamefont {Lopez}(2022)}]{Lopez22t}%
  \BibitemOpen
  \bibfield  {author} {\bibinfo {author} {\bibfnamefont {N.~A.}\ \bibnamefont {Lopez}},\ }\emph {\bibinfo {title} {Metaplectic geometrical optics}},\ \href {https://arxiv.org/abs/2210.03188} {Ph.D. thesis},\ \bibinfo  {school} {Princeton University} (\bibinfo {year} {2022})\BibitemShut {NoStop}%
\bibitem [{\citenamefont {Dixit}\ \emph {et~al.}(1993)\citenamefont {Dixit}, \citenamefont {Thomas}, \citenamefont {Woods}, \citenamefont {Morgan}, \citenamefont {Henesian}, \citenamefont {Wegner},\ and\ \citenamefont {Powell}}]{Dixit93}%
  \BibitemOpen
  \bibfield  {author} {\bibinfo {author} {\bibfnamefont {S.~N.}\ \bibnamefont {Dixit}}, \bibinfo {author} {\bibfnamefont {I.~M.}\ \bibnamefont {Thomas}}, \bibinfo {author} {\bibfnamefont {B.~W.}\ \bibnamefont {Woods}}, \bibinfo {author} {\bibfnamefont {A.~J.}\ \bibnamefont {Morgan}}, \bibinfo {author} {\bibfnamefont {M.~A.}\ \bibnamefont {Henesian}}, \bibinfo {author} {\bibfnamefont {P.~J.}\ \bibnamefont {Wegner}}, \ and\ \bibinfo {author} {\bibfnamefont {H.~T.}\ \bibnamefont {Powell}},\ }\href {\doibase 10.1364/AO.32.002543} {\bibfield  {journal} {\bibinfo  {journal} {Appl. Opt.}\ }\textbf {\bibinfo {volume} {32}},\ \bibinfo {pages} {2543} (\bibinfo {year} {1993})}\BibitemShut {NoStop}%
\bibitem [{\citenamefont {Lopez}\ and\ \citenamefont {Dodin}(2020)}]{Lopez20}%
  \BibitemOpen
  \bibfield  {author} {\bibinfo {author} {\bibfnamefont {N.~A.}\ \bibnamefont {Lopez}}\ and\ \bibinfo {author} {\bibfnamefont {I.~Y.}\ \bibnamefont {Dodin}},\ }\href {\doibase 10.1088/1367-2630/aba91a} {\bibfield  {journal} {\bibinfo  {journal} {New J. Phys.}\ }\textbf {\bibinfo {volume} {22}},\ \bibinfo {pages} {083078} (\bibinfo {year} {2020})}\BibitemShut {NoStop}%
\bibitem [{\citenamefont {Lopez}\ and\ \citenamefont {Dodin}(2021)}]{Lopez21a}%
  \BibitemOpen
  \bibfield  {author} {\bibinfo {author} {\bibfnamefont {N.~A.}\ \bibnamefont {Lopez}}\ and\ \bibinfo {author} {\bibfnamefont {I.~Y.}\ \bibnamefont {Dodin}},\ }\href {\doibase 10.1088/2040-8986/abd1ce} {\bibfield  {journal} {\bibinfo  {journal} {J. Opt.}\ }\textbf {\bibinfo {volume} {23}},\ \bibinfo {pages} {025601} (\bibinfo {year} {2021})}\BibitemShut {NoStop}%
\bibitem [{\citenamefont {Lopez}\ and\ \citenamefont {Dodin}(2022)}]{Lopez22}%
  \BibitemOpen
  \bibfield  {author} {\bibinfo {author} {\bibfnamefont {N.~A.}\ \bibnamefont {Lopez}}\ and\ \bibinfo {author} {\bibfnamefont {I.~Y.}\ \bibnamefont {Dodin}},\ }\href {\doibase 10.1063/5.0082241} {\bibfield  {journal} {\bibinfo  {journal} {Phys. Plasmas}\ }\textbf {\bibinfo {volume} {29}},\ \bibinfo {pages} {052111} (\bibinfo {year} {2022})}\BibitemShut {NoStop}%
\bibitem [{\citenamefont {Lopez}\ \emph {et~al.}(2024)\citenamefont {Lopez}, \citenamefont {{H\o jlund}},\ and\ \citenamefont {Senstius}}]{Lopez24b}%
  \BibitemOpen
  \bibfield  {author} {\bibinfo {author} {\bibfnamefont {N.~A.}\ \bibnamefont {Lopez}}, \bibinfo {author} {\bibfnamefont {R.}~\bibnamefont {{H\o jlund}}}, \ and\ \bibinfo {author} {\bibfnamefont {M.~G.}\ \bibnamefont {Senstius}},\ }\href {https://arxiv.org/abs/2406.01270} {\bibfield  {journal} {\bibinfo  {journal} {arXiv:2406.01270}\ } (\bibinfo {year} {2024})}\BibitemShut {NoStop}%
\bibitem [{\citenamefont {{H\o jlund}}\ \emph {et~al.}(2024)\citenamefont {{H\o jlund}}, \citenamefont {Senstius},\ and\ \citenamefont {Nielsen}}]{Hojlund24}%
  \BibitemOpen
  \bibfield  {author} {\bibinfo {author} {\bibfnamefont {R.}~\bibnamefont {{H\o jlund}}}, \bibinfo {author} {\bibfnamefont {M.~G.}\ \bibnamefont {Senstius}}, \ and\ \bibinfo {author} {\bibfnamefont {S.~K.}\ \bibnamefont {Nielsen}},\ }\href {https://arxiv.org/abs/2402.03882} {\bibfield  {journal} {\bibinfo  {journal} {arXiv:2402.03882}\ } (\bibinfo {year} {2024})}\BibitemShut {NoStop}%
\bibitem [{\citenamefont {Lopez}\ \emph {et~al.}(2021)\citenamefont {Lopez}, \citenamefont {Kur}, \citenamefont {Chapman}, \citenamefont {Strozzi},\ and\ \citenamefont {Michel}}]{Lopez21DPP}%
  \BibitemOpen
  \bibfield  {author} {\bibinfo {author} {\bibfnamefont {N.~A.}\ \bibnamefont {Lopez}}, \bibinfo {author} {\bibfnamefont {E.}~\bibnamefont {Kur}}, \bibinfo {author} {\bibfnamefont {T.~D.}\ \bibnamefont {Chapman}}, \bibinfo {author} {\bibfnamefont {D.~J.}\ \bibnamefont {Strozzi}}, \ and\ \bibinfo {author} {\bibfnamefont {P.~A.}\ \bibnamefont {Michel}},\ }\href {\doibase https://meetings.aps.org/Meeting/DPP21/Session/NP11.63} {\bibfield  {journal} {\bibinfo  {journal} {Bull. Am. Phys. Soc.}\ }\textbf {\bibinfo {volume} {66}},\ \bibinfo {pages} {Abstract NP11.00063} (\bibinfo {year} {2021})}\BibitemShut {NoStop}%
\bibitem [{\citenamefont {Spaeth}\ \emph {et~al.}(2016)\citenamefont {Spaeth}, \citenamefont {Manes}, \citenamefont {Kalantar}, \citenamefont {Miller}, \citenamefont {Heebner}, \citenamefont {Bliss}, \citenamefont {Spec}, \citenamefont {Parham}, \citenamefont {Whitman}, \citenamefont {Wegner}, \citenamefont {Baisden}, \citenamefont {Menapace}, \citenamefont {Bowers}, \citenamefont {Cohen}, \citenamefont {Suratwala}, \citenamefont {{Di Nicola}}, \citenamefont {Newton}, \citenamefont {Adams}, \citenamefont {Trenholme}, \citenamefont {Finucane}, \citenamefont {Bonanno}, \citenamefont {Rardin}, \citenamefont {Arnold}, \citenamefont {Dixit}, \citenamefont {Erbert}, \citenamefont {Erlandson}, \citenamefont {Fair}, \citenamefont {Feigenbaum}, \citenamefont {Gourdin}, \citenamefont {Hawley}, \citenamefont {Honig}, \citenamefont {House}, \citenamefont {Jancaitis}, \citenamefont {LaFortune}, \citenamefont {Larson}, \citenamefont {{Le Galloudec}}, \citenamefont {Lindl}, \citenamefont {MacGowan}, \citenamefont
  {Marshall}, \citenamefont {McCandless}, \citenamefont {McCracken}, \citenamefont {Montesanti}, \citenamefont {Moses}, \citenamefont {Nostrand}, \citenamefont {Pryatel}, \citenamefont {Roberts}, \citenamefont {Rodrigues}, \citenamefont {Rowe}, \citenamefont {Sacks}, \citenamefont {Salmon}, \citenamefont {Shaw}, \citenamefont {Sommer}, \citenamefont {Stolz}, \citenamefont {Tietbohl}, \citenamefont {Widmayer},\ and\ \citenamefont {Zacharias}}]{Spaeth16}%
  \BibitemOpen
  \bibfield  {author} {\bibinfo {author} {\bibfnamefont {M.~L.}\ \bibnamefont {Spaeth}}, \bibinfo {author} {\bibfnamefont {K.~R.}\ \bibnamefont {Manes}}, \bibinfo {author} {\bibfnamefont {D.~G.}\ \bibnamefont {Kalantar}}, \bibinfo {author} {\bibfnamefont {P.~E.}\ \bibnamefont {Miller}}, \bibinfo {author} {\bibfnamefont {J.~E.}\ \bibnamefont {Heebner}}, \bibinfo {author} {\bibfnamefont {E.~S.}\ \bibnamefont {Bliss}}, \bibinfo {author} {\bibfnamefont {D.~R.}\ \bibnamefont {Spec}}, \bibinfo {author} {\bibfnamefont {T.~G.}\ \bibnamefont {Parham}}, \bibinfo {author} {\bibfnamefont {P.~K.}\ \bibnamefont {Whitman}}, \bibinfo {author} {\bibfnamefont {P.~J.}\ \bibnamefont {Wegner}}, \bibinfo {author} {\bibfnamefont {P.~A.}\ \bibnamefont {Baisden}}, \bibinfo {author} {\bibfnamefont {J.~A.}\ \bibnamefont {Menapace}}, \bibinfo {author} {\bibfnamefont {M.~W.}\ \bibnamefont {Bowers}}, \bibinfo {author} {\bibfnamefont {S.~J.}\ \bibnamefont {Cohen}}, \bibinfo {author} {\bibfnamefont {T.~I.}\ \bibnamefont {Suratwala}},
  \bibinfo {author} {\bibfnamefont {J.~M.}\ \bibnamefont {{Di Nicola}}}, \bibinfo {author} {\bibfnamefont {M.~A.}\ \bibnamefont {Newton}}, \bibinfo {author} {\bibfnamefont {J.~J.}\ \bibnamefont {Adams}}, \bibinfo {author} {\bibfnamefont {J.~B.}\ \bibnamefont {Trenholme}}, \bibinfo {author} {\bibfnamefont {R.~G.}\ \bibnamefont {Finucane}}, \bibinfo {author} {\bibfnamefont {R.~E.}\ \bibnamefont {Bonanno}}, \bibinfo {author} {\bibfnamefont {D.~C.}\ \bibnamefont {Rardin}}, \bibinfo {author} {\bibfnamefont {P.~A.}\ \bibnamefont {Arnold}}, \bibinfo {author} {\bibfnamefont {S.~N.}\ \bibnamefont {Dixit}}, \bibinfo {author} {\bibfnamefont {G.~V.}\ \bibnamefont {Erbert}}, \bibinfo {author} {\bibfnamefont {A.~C.}\ \bibnamefont {Erlandson}}, \bibinfo {author} {\bibfnamefont {J.~E.}\ \bibnamefont {Fair}}, \bibinfo {author} {\bibfnamefont {E.}~\bibnamefont {Feigenbaum}}, \bibinfo {author} {\bibfnamefont {W.~H.}\ \bibnamefont {Gourdin}}, \bibinfo {author} {\bibfnamefont {R.~A.}\ \bibnamefont {Hawley}}, \bibinfo {author}
  {\bibfnamefont {J.}~\bibnamefont {Honig}}, \bibinfo {author} {\bibfnamefont {R.~K.}\ \bibnamefont {House}}, \bibinfo {author} {\bibfnamefont {K.~S.}\ \bibnamefont {Jancaitis}}, \bibinfo {author} {\bibfnamefont {K.~N.}\ \bibnamefont {LaFortune}}, \bibinfo {author} {\bibfnamefont {D.~W.}\ \bibnamefont {Larson}}, \bibinfo {author} {\bibfnamefont {B.~J.}\ \bibnamefont {{Le Galloudec}}}, \bibinfo {author} {\bibfnamefont {J.~D.}\ \bibnamefont {Lindl}}, \bibinfo {author} {\bibfnamefont {B.~J.}\ \bibnamefont {MacGowan}}, \bibinfo {author} {\bibfnamefont {C.~D.}\ \bibnamefont {Marshall}}, \bibinfo {author} {\bibfnamefont {K.~P.}\ \bibnamefont {McCandless}}, \bibinfo {author} {\bibfnamefont {R.~W.}\ \bibnamefont {McCracken}}, \bibinfo {author} {\bibfnamefont {R.~C.}\ \bibnamefont {Montesanti}}, \bibinfo {author} {\bibfnamefont {E.~I.}\ \bibnamefont {Moses}}, \bibinfo {author} {\bibfnamefont {M.~C.}\ \bibnamefont {Nostrand}}, \bibinfo {author} {\bibfnamefont {J.~A.}\ \bibnamefont {Pryatel}}, \bibinfo {author}
  {\bibfnamefont {V.~S.}\ \bibnamefont {Roberts}}, \bibinfo {author} {\bibfnamefont {S.~B.}\ \bibnamefont {Rodrigues}}, \bibinfo {author} {\bibfnamefont {A.~W.}\ \bibnamefont {Rowe}}, \bibinfo {author} {\bibfnamefont {R.~A.}\ \bibnamefont {Sacks}}, \bibinfo {author} {\bibfnamefont {J.~T.}\ \bibnamefont {Salmon}}, \bibinfo {author} {\bibfnamefont {M.~J.}\ \bibnamefont {Shaw}}, \bibinfo {author} {\bibfnamefont {S.}~\bibnamefont {Sommer}}, \bibinfo {author} {\bibfnamefont {C.~J.}\ \bibnamefont {Stolz}}, \bibinfo {author} {\bibfnamefont {G.~L.}\ \bibnamefont {Tietbohl}}, \bibinfo {author} {\bibfnamefont {C.~C.}\ \bibnamefont {Widmayer}}, \ and\ \bibinfo {author} {\bibfnamefont {R.}~\bibnamefont {Zacharias}},\ }\href {\doibase 10.13182/FST15-144} {\bibfield  {journal} {\bibinfo  {journal} {Fusion Sci. Technol.}\ }\textbf {\bibinfo {volume} {69}},\ \bibinfo {pages} {25} (\bibinfo {year} {2016})}\BibitemShut {NoStop}%
\bibitem [{\citenamefont {Gil}\ \emph {et~al.}(2003)\citenamefont {Gil}, \citenamefont {Segura},\ and\ \citenamefont {Temme}}]{Gil03}%
  \BibitemOpen
  \bibfield  {author} {\bibinfo {author} {\bibfnamefont {A.}~\bibnamefont {Gil}}, \bibinfo {author} {\bibfnamefont {J.}~\bibnamefont {Segura}}, \ and\ \bibinfo {author} {\bibfnamefont {N.~M.}\ \bibnamefont {Temme}},\ }\href {\doibase 10.1016/S0021-9045(02)00022-9} {\bibfield  {journal} {\bibinfo  {journal} {J. Approx. Theory}\ }\textbf {\bibinfo {volume} {120}},\ \bibinfo {pages} {253} (\bibinfo {year} {2003})}\BibitemShut {NoStop}%
\end{thebibliography}%
\bibliographystyle{apsrev4-1}

\end{document}